\documentstyle[pra,aps,tighten,eqsecnum,epsfig]{revtex}

\title{Disagreement between correlations of quantum mechanics and stochastic 
electrodynamics in the damped parametric oscillator}
\author{D. T. Pope, P. D. Drummond \& W. J. Munro \\
Centre for Laser Science and Department of Physics, University of Queensland,
Brisbane 4072, Queensland, Australia \\
(pope@physics.uq.edu.au)}
\date{\today}

\begin{document}

\maketitle

\begin{abstract}
Intracavity and external third order correlations in the damped nondegenerate 
parametric oscillator are calculated for
quantum mechanics and stochastic electrodynamics (SED), a semiclassical theory. 
The two theories yield greatly different results,
with the correlations of quantum mechanics being cubic in the system's 
nonlinear
coupling constant and those of SED being linear in the same constant. 
In particular, 
differences between the two theories are present in at least a mesoscopic 
regime. They also
exist when realistic damping is included. Such differences illustrate
distinctions between quantum mechanics and a hidden variable theory for 
continuous variables.
\end{abstract}


\section{Introduction}

Local hidden variable theories have been extensively compared to
quantum mechanics over the last seventy or so years
\cite{epr,bell,general}. Most comparisons between the two have investigated
whether or not quantum mechanics is equivalent to a local hidden
variable theory. Much evidence indicates that it is not. Many 
results in quantum mechanics have been found that are incompatible with all 
local hidden 
variable theories \cite{bell,general,ghz,damp_bell,large,continuous}.
Most of these results have involved 
idealized undamped systems. However, all experimental systems encounter damping.
Thus, it is interesting (and more realistic) to compare quantum mechanics
and local hidden variable theories in damped systems \cite{damp_bell}. 
This paper compares quantum mechanics and one local hidden variable theory (SED)
in such a system.

One of the earliest works comparing local hidden variable theories to quantum 
mechanics was
Bell's theorem \cite{bell}. It demonstrates that quantum
mechanics is incompatible with all local hidden variable theories at a 
statistical 
level. It does so by deriving an upper bound on a function of two particle
correlations for all local hidden variable theories, which quantum mechanics 
exceeds. 
Extensions of it have been formulated for
large angular momentum and
particle number systems \cite{damp_bell,large}. These extensions demonstrate 
nonclassical  
behavior in a regime usually regarded as being purely classical. 
Greenberger, Horne and Zeilinger (GHZ) \cite{ghz} have also extended Bell's 
work, differentiating quantum mechanics from all local hidden variable theories
for single, as opposed to ensemble, measurements. The three particle GHZ 
theorem has an ``all or nothing"
quality and distinguishes between local hidden variable theories and quantum
mechanics in a single experimental run, once three basic correlations
are established.

Comparisons between quantum mechanics and local hidden variable theories
have also been made using continuous variables (which are discretized
in formulating the comparison), such as
quadrature phase amplitudes \cite{continuous} and there is currently much
interest in this area. For quadrature phase amplitude measurements, these 
comparisons can 
have detector efficiencies of in excess of 99\% \cite{efficient}. They also 
tend to relate more
strongly to Einstein, Podolsky and Rosen's original EPR paradox \cite{epr} than 
earlier discrete variable ones. Indeed, the EPR paradox has been experimentally demonstrated
using quadrature phase amplitudes \cite{expt_epr}.
Additionally, quantum teleportation has been achieved using quadrature phase 
amplitudes
\cite{teleport} further demonstrating the utility of continuous variables.

One commonly used local hidden variable theory is
stochastic electrodynamics (SED) \cite{marshall,boyer}. Some authors have
proposed it as an 
alternative to quantum mechanics \cite{marshall,alternative}. 
Furthermore, a semiclassical approach equivalent to it \cite{semi_method} 
is also commonly used in parametric oscillator calculations \cite{common}.
SED consists of adding Gaussian white noise to classical electrodynamics.
It is equivalent to truncating third order derivative
terms in the quantum mechanical Moyal equation, a commonly used
approximation \cite{graham}. Such terms are often 
negligible and thus SED reproduces many results of quantum mechanics
\cite{semi_method,common}. 
However, it cannot violate Bell inequalities for quadrature phase amplitude
measurements, and is thus distinct 
from quantum mechanics \cite{continuous}. Various authors have explicitly shown 
differences
between SED and quantum mechanics \cite{drummond,kinsler,new_compare}.
In particular, it has been shown that the two theories predict different 
transient third order correlations for the undamped nondegenerate parametric 
oscillator \cite{drummond}. It has also been shown that they 
predict different macroscopic quadrature phase amplitude correlations in the 
damped nondegenerate parametric oscillator in the steady state 
\cite{new_compare}.

In general, differences between quantum mechanics and local hidden variable 
theories are reduced or eliminated by damping \cite{decrease}. 
Furthermore, damping is a significant 
element of many realistic systems. It is thus important to consider its
effects on differences between quantum mechanics and local hidden variable
theories such as SED. However, all but a few of the comparisons 
between the quantum mechanics and local hidden variable theories referenced 
above have involved undamped systems. They are thus idealized in this respect. 
In contrast, damping is included in the calculatons in this 
paper. It is included to consider a theoretical model which is as realistic  
as possible and also to determine the sensitivity of 
differences between quantum mechanics and SED to its presence.

This paper extends a previous comparison 
between quantum mechanics and SED in the nondegenerate parametric 
oscillator \cite{drummond}. In particular, it contrasts both  
intracavity and external moments of the two theories' in the same system with 
damping 
included. Expressions from both theories are compared 
for the intracavity moment $\langle \Delta X_{1}(\tau) \Delta (X_{2}(\tau) 
\Delta X_{3}(\tau) \rangle$, 
where $\Delta X_{i}(\tau)=X_{i}(\tau)-\langle X_{i}(\tau) \rangle$, for
$i=1,2,3$, $X_{i}(\tau)$ is quadrature phase amplitude, the subscripts 
represent
different radiation modes and $\tau$ is a scaled time variable. 
A comparison is also made for an analogous external moment.
Both analytic iterative and numerical techniques 
are used to calculate moments. The results produced by these techniques show
that the intracavity and external moments differ greatly 
between the two theories. 
In particular, the analytic method shows that the
moments of quantum mechanics are cubic in the system's nonlinear coupling
constant to leading order whilst those of SED 
are linear.
The two theories are compared over a range of nonlinear coupling constant,
damping and average initial pump photon number values. The results 
of these comparisons show a number of 
qualitative trends. Most importantly, quantum mechanics and SED differ in the 
situations 
considered with the largest
particle number and damping to nonlinear coupling ratios, although the 
differences are reduced in 
relative size. 

Stochastic techniques are used to obtain results both for quantum
mechanics and SED. The positive-P coherent state representation 
\cite{positive_p}
is used to calculate quantum mechanical predictions. It is particularly well
suited to the calculation of quantum dynamics in damped quantum optical systems
when nonclassical behavior is present. It is able to handle arbitrarily
large photon numbers. It converges quickly (in the sense of sampling error) when
systems' dimensionless nonlinearities are relatively small, as is the case with
nonlinear optical experiments. By contrast, the method used for SED 
calculations corresponds to
commonly used approaches in quantum optics, where the field is treated as a 
semiclassical object surrounded by (classical) vacuum fluctuations. Both methods
are used to generate analytic predictions
and are also numerically simulated.

\section{Quantum mechanics} \label{section_two}

This paper considers an idealized nondegenerate parametric oscillator, resonant
at three frequencies $\omega_{1}\:{\rm and}\:\:\:\omega_{2}$ (signal and idler
frequencies) and $\omega_{3}\:=\omega_{1}+\omega_{2}$ (pump 
frequency). It contains a nonlinear medium that couples the modes and converts
higher energy pump photons into lower energy signal and idler ones.
The system's interaction Hamiltonian, including linear losses, is given by

\begin{equation}
\hat{H}=i\hbar G({\hat{a}}^{\dag}_{1} {\hat{a}}^{\dag}_{2}{\hat{a}}_{3}-
{\hat{a}}_{1}{\hat{a}}_{2}{\hat{a}}^{\dag}_{3})
+\sum^{3}_{i=1} {\hat{\Gamma}}_{i} {\hat{a}}_{i}^{\dag} + 
{\hat{\Gamma}}_{i}^{\dag} {\hat{a}}_{i},
\end{equation}
where ${\hat{a}}_{i}^{\dag}\:{\rm and}\:{\hat{a}}_{i}$ are creation and 
annihilation operators
for oscillator modes, ${\hat{\Gamma}}^{\dag}_{i}\:{\rm and}\:{\hat{\Gamma}}_{i}$ 
are 
environment mode operators and G is a nonlinear 
interaction strength constant.
Initially, the system has a coherent state in the pump mode, and 
vacuum states in
the signal and idler modes.

A number of quasiprobability representations exist to describe
quantum states, the most famous being the Glauber-Sudarshan
representation\cite{Glauber}. It is produced by decomposing quantum density 
operators using
a diagonal coherent state basis. Thus, 
\begin{equation}
{\hat \rho}=\int d\alpha^{2} P(\alpha,\alpha^{*}) | \alpha \rangle
\langle \alpha |,
\end{equation}
where ${\hat \rho}$ is a density operator and $P(\alpha,\alpha^{*})$
is the Glauber-Sudarshan representation.
The Glauber-Sudarshan representation
can be negative and is hence not a strict probability
density function. A more recent representation is the
positive-P representation \cite{positive_p} which  
is an actual 
probability density function over an off-diagonal coherent state basis. 
It further differs from the Glauber-Sudarshan representation by using a phase
space of doubled dimension.
The positive-P variables
$ \{ \alpha_{i},\alpha^{+}_{i}\} $, where $i$ is a positive integer, are 
analogous to complex field
amplitudes, with $\alpha_{i}$ and
$\alpha^{+}_{i}$ describing a particular radiation mode.  
However, $\{\alpha_{i}\}$ and $\{\alpha^{+}_{i}\}$ are independent and hence
$\alpha_{i}\neq (\alpha^{+}_{i})^{*}$, though their averages are complex
conjugate and thus $\langle \alpha_{i} \rangle=\langle \alpha_{i}^{+} 
\rangle^{*}$.
Variable averages are equal to normally ordered quantum averages once
the substitutions
$\alpha_{i} \rightarrow {\hat{a}}_{i}$ and $\alpha_{i}^{+} \rightarrow
{\hat{a}}_{i}^{\dag}$
are made.
For example, $\langle \alpha_{1} \alpha_{1}^{+} \rangle=\langle 
{\hat{a}}_{1}^{\dag}
{\hat{a}}_{1}\rangle_{\rho}$, where $\langle \hat{O} \rangle_{\rho}$
denotes $Tr(\hat{\rho}\hat{O})$, as usual in quantum mechanics.

Stochastic equations of motion for positive-P variables for the damped 
nondegenerate
parametric oscillator are, in terms of $\tau$ (time scaled  by $\Gamma$, a
typical damping constant with units of inverse time),
\begin{eqnarray} \label{positive_p}
\frac{d\alpha_{1}}{d\tau} & =  & - \gamma_{1} \alpha_{1} + g\alpha_{2}^{+} 
\alpha_{3}
+\sqrt{g\alpha_{3}} \xi_{1} \nonumber \\
\frac{d\alpha_{1}^{+}}{d\tau} & =  & - \gamma_{1} \alpha_{1}^{+} + 
g\alpha_{2}\alpha_{3}^{+} + \sqrt{g\alpha_{3}^{+}} \xi_{1}^{+} \nonumber \\
\frac{d\alpha_{2}}{d\tau} & =  & - \gamma_{2} \alpha_{2} + g\alpha_{1}^{+} 
\alpha_{3}
+\sqrt{g\alpha_{3}} \xi_{2} \nonumber \\
\frac{d\alpha_{2}^{+}}{d\tau} & =  & - \gamma_{2} \alpha_{2}^{+} + 
g\alpha_{1}\alpha_{3}^{+} + \sqrt{g\alpha_{3}^{+}} \xi_{2}^{+}
\nonumber \\
\frac{d\alpha_{3}}{d\tau} & =  & - \gamma_{3} \alpha_{3} - g\alpha_{1} 
\alpha_{2} 
\nonumber \\
\frac{d\alpha_{3}^{+}}{d\tau} & =  & - \gamma_{3} \alpha_{3}^{+} - 
g\alpha_{1}^{+}\alpha_{2}^{+}. 
\end{eqnarray}
Here $\xi_{1},\:\xi_{2},\:\xi^{+}_{1} {\rm \:and\:} \xi^{+}_{2}$ are complex
Gaussian white noises with the following correlations 
\begin{eqnarray}
\langle \xi_{i}(\tau_{1}) \xi_{j}(\tau_{2})\rangle&=&
\delta_{3-i,j} \delta(\tau_{1}-\tau_{2}) \\ \nonumber  
\langle \xi_{i}^{+}(\tau_{1}) \xi_{j}^{+}(\tau_{2})\rangle&=&
\delta_{3-i,j} \delta(\tau_{1}-\tau_{2}) \\ \nonumber
\langle \xi_{i}^{+}(\tau_{1})\xi_{j}(\tau_{2})\rangle&=&0,
\end{eqnarray}
where $i,j=1,2$. In Eq.~(\ref{positive_p}), $\gamma_{i}=\Gamma_{i}/\Gamma$, 
where $\Gamma_{i}$ is a damping constant for mode $i$ with units of inverse
time, $g=G / \Gamma$ and $\tau=\Gamma t$. It is assumed that G, $\Gamma_{i}$ 
and $\Gamma$ are real.
Initial conditions are $\alpha_{1}(0)=0,\:\alpha_{2}(0)=0\:{\rm
and}\:\alpha_{3}(0)=\epsilon$. It is noted that Eq.~(\ref{positive_p})
is only valid when boundary terms in phase space can be neglected. These 
are asymptotically small in the limit of short times or large damping ratios
\cite{neglect}.

Eq.~\ref{positive_p} is solved using an analytic iterative method.
This method treats damping terms
exactly, and noise and nonlinear terms iteratively.
It involves, firstly, rewriting the equations forming 
Eq.~(\ref{positive_p}) as 
$\dot \alpha_{i}=-\gamma_{i}\alpha_{i} + 
f_{i}(\{\alpha_{j},\alpha^{+}_{j}\},\tau)$
or $\dot \alpha_{i}^{+}=-\gamma_{i} \alpha_{i}^{+} + 
f^{+}_{i}(\{\alpha_{j},\alpha^{+}_{j}\},\tau)$, where $i,j=1,2,3$. 
Successively higher order approximations for 
$\{\alpha_{i}(\tau),\alpha_{i}^{+}(\tau)\}$ are then found using increasingly
better approximations for $f_{i}$ and $f_{i}^{+}$.
Thus, $(m+1)^{th}$ order terms are given by

\begin{eqnarray}
\alpha_{i}^{(m+1)}(\tau)&=&\alpha_{i}^{(0)}(\tau)+\int_{\tau_{1}=0}^
{\tau_{1}=\tau}d\tau_{1}\exp[{\gamma_{i}(\tau_{1}-\tau)}]
f_{i}(\{\alpha^{(m)}_{j},\alpha^{+\:(m)}_{j}\},\tau_{1}) \\ \nonumber
\alpha_{i}^{+\:(m+1)}(\tau)&=&\alpha_{i}^{+\:(0)}(\tau)+
\int_{\tau_{1}=0}^{\tau_{1}=\tau}d\tau_{1}\exp[{\gamma_{i}(\tau_{1}-\tau)}]
f^{+}_{i}(\{\alpha^{(m)}_{j},\alpha^{+\:(m)}_{j}\},\tau_{1}),
\label{iterate}
\end{eqnarray}
where $\alpha_{k}^{(0)}(\tau)=\alpha_{k}(\tau=0)\exp(-\gamma_{k}\tau)$
and $\alpha_{k}^{(0)\:+}(\tau)=\alpha_{k}^{(0)}(\tau)^{*}$ where $k=1,2,3$.
For example,
\begin{equation}
\alpha_{1}^{(m+1)}(\tau)=\alpha_{1}^{(0)}(\tau)+
\int_{\tau_{1}=0}^{\tau_{1}=\tau}d\tau_{1}\exp[{\gamma_{1}(\tau_{1}-\tau)}]
\bigl( g \alpha_{2}^{+\:(m)}(\tau_{1})
\alpha_{3}^{(m)}(\tau_{1})+\sqrt{g\alpha_{3}^{(m)}(\tau_{1})}
\xi_{1}(\tau_{1})\bigr)
\end{equation}
and first order approximations are 
\begin{eqnarray}
\alpha_{i}^{(1)}(\tau)&=& \int_{\tau_{i}=0}^{\tau_{i}=\tau}
d\tau_{i}\exp[\gamma_{i}(\tau_{i}-\tau)]\sqrt{g\epsilon}
\exp(-\frac{\gamma_{i}\tau_{i}}{2})
\xi_{i} (\tau_{1}) \\ \nonumber
\alpha_{3}^{(1)}(\tau)&=&\epsilon \exp(-\gamma_{3}\tau) \\ \nonumber
\alpha_{i}^{+\:(1)}(\tau)&=& \int_{\tau_{i}=0}^{\tau_{i}=\tau}d\tau_{i}
\exp[\gamma_{i}(\tau_{i}-\tau)]\sqrt{g\epsilon^{*}}
\exp(-\frac{\gamma_{i}\tau_{i}}{2})
\xi_{i}^{+} (\tau_{1}) \\ \nonumber
\alpha_{3}^{+\:(1)}(\tau)&=&\epsilon^{*} \exp(-\gamma_{3}\tau),
\end{eqnarray} 
where $i=1,2$.

\section{Stochastic diagrams} \label{stoch_section}

\begin{figure}
\center{\epsfig{figure=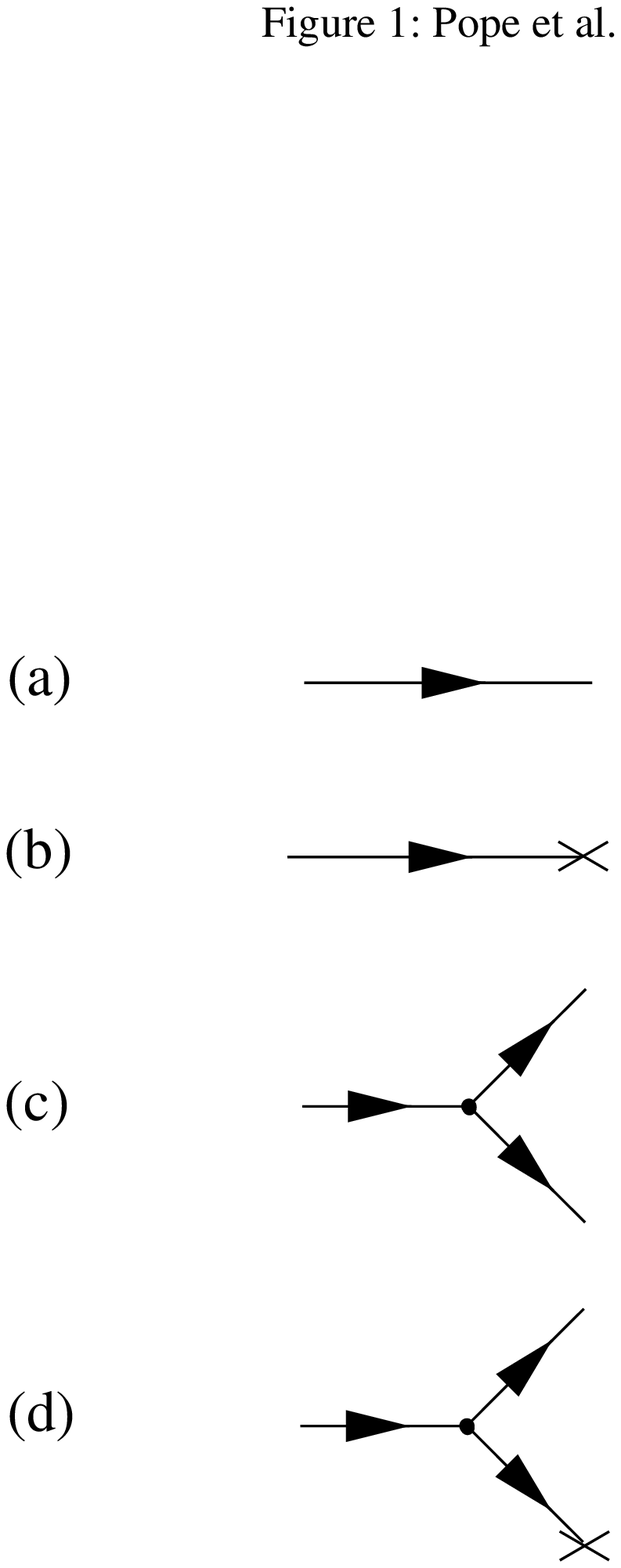,width=75mm}} 
\caption{The three basic classes of stochastic diagrams, (a) initital value
term, (b) noise term and (c) nonlinear term. A stochastic diagram (d)
representing a higher order term.}
\label{fig1}
\end{figure}

The iterative method of the previous section can be used, in conjunction with
stochastic diagrams, to readily produce analytic approximations for the
intracavity moments of quantum mechanics considered in this paper.
Stochastic diagrams \cite{stoch_diag} are schematic representations of the 
combinatoric parts of an
iterative process. They clearly lay out all terms produced by different orders
of iteration. Fundamental stochastic diagrams appear as one of three classes.
Those associated with initial conditions appear as straight
lines, those with noise terms as straight lines with a cross at their end and
those with nonlinear terms as straight lines containing a fork, as shown in 
Fig.~\ref{fig1}(a)-(c). 
Higher order iterative terms are represented by
stochastic diagrams using either combinations of the three basic classes. 
For example, one of the iterative terms in
$\alpha_{1}^{(2)}(\tau)$ is 
\begin{displaymath}
\int_{\tau_{1}=0}^{\tau_{1}=\tau}d\tau_{1}\exp[\gamma_{1}(\tau_{1}-\tau)]
g\alpha_{3}^{(0)}(\tau_{1})\int_{\tau_{2}=0}^{\tau_{2}=\tau_{1}}d\tau_{2}
\exp[\gamma_{2}(\tau_{2}-\tau_{1})]\sqrt{g\epsilon^{*}}
\exp(-\frac{\gamma_{2}\tau_{2}}{2})\xi_{2}^{+} (\tau_{2}).
\end{displaymath}
It combines all three basic classes and is represented by
the stochastic diagram in Fig.~\ref{fig1}(d). All iterative terms can be
represented by stochastic diagrams.

Stochastic diagrams can also be used to determine the orders of iterative terms.
In particular, they can be used to determine the orders of such terms in
the system's nonlinear coupling constant g. This paper focuses on the order of 
terms in this constant. For quantum mechanics, initial value iterative terms
are $O(g^{0})$, noise iterative terms $O(g^{1/2})$ and nonlinear iterative terms
$O(g)$. Hence, lines in stochastic diagrams count as order zero, crosses as
order 1/2 and vertices as order 1. A term's order is simply found by considering
its stochastic diagram and adding a half to its order for every cross and one
for every vertex.
For example, the term represented in Fig.~\ref{fig1}(d) has one vertex
and one cross and thus is $O(g^{\frac{3}{2}})$. A notation that denotes the 
order in g of a term by a superscript [n] is used in this section.

Stochastic diagrams are now used to determine the intracavity moments
of quantum mechanics considered in this paper.
Consider all eight moments of the form
$\langle \Delta {\cal A}_{1} (\tau) \Delta  {\cal A}_{2}(\tau) 
\Delta {\cal A}_{3} (\tau) \rangle$, 
where $\Delta {\rm {\cal A}_{i}}(\tau)={\cal A}_{i}(\tau)-
\langle {\cal A}_{i}(\tau)\rangle$  and
${\cal A}_{i}(\tau)$ is either ${\hat a}_{i}$ or ${\hat a}^{\dag}_{i}$. 
These are equal to the positive-P variable moments
which replace ${\hat a}_{i}$ and $\hat{a}_{i}^{\dag}$ by 
$\alpha_{i}(\tau)$ and $\alpha_{i}^{+}(\tau)$ respectively.
Now, consider the equations that constitute  Eq.~(\ref{positive_p}).
Their forms do not change when they are expressed
in terms of $-\alpha_{i}(\tau),-\alpha_{i}^{+}(\tau),\alpha_{3}(\tau)$ and 
$\alpha_{3}^{+}(\tau)$,
where $i=1,2$.
From this, it follows that $\langle {\cal A}_{i}(\tau) \rangle = \langle - 
{\cal A}_{i}
(\tau) \rangle$
and hence $\langle \alpha_{i}(\tau) \rangle= \langle \alpha_{i}^{+}(\tau) 
\rangle=0$,
where again $i=1,2$.
Thus, $\langle \Delta {\cal A}_{1}(\tau) \Delta  {\cal A}_{2}(\tau) 
\Delta {\cal A}_{3}(\tau) \rangle$
can be simplified to 
$\langle {\cal A}_{1} (\tau){\cal A}_{2}(\tau) \Delta {\cal A}_{3}(\tau) 
\rangle$.

An approximate expression for $\langle {\cal A}_{1} (\tau){\cal A}_{2}(\tau) 
\Delta {\cal A}_{3}(\tau) \rangle$ is now obtained using the iterative method
in Section \ref{section_two} (and stochastic diagrams). This method can be used to produce power series 
expressions in $g$ for the positive-P variables.
These expressions can then be used to 
generate power series expressions in $g$ for the moments of the form 
$\langle {\cal A}_{1}(\tau){\cal A}_{2}(\tau)\Delta {\cal A}_{3}(\tau)\rangle$.
As $g \ll 1$ in realistic systems, these power series expressions can be
approximated by their lowest order nonzero terms.

\begin{figure}
\center{\epsfig{figure=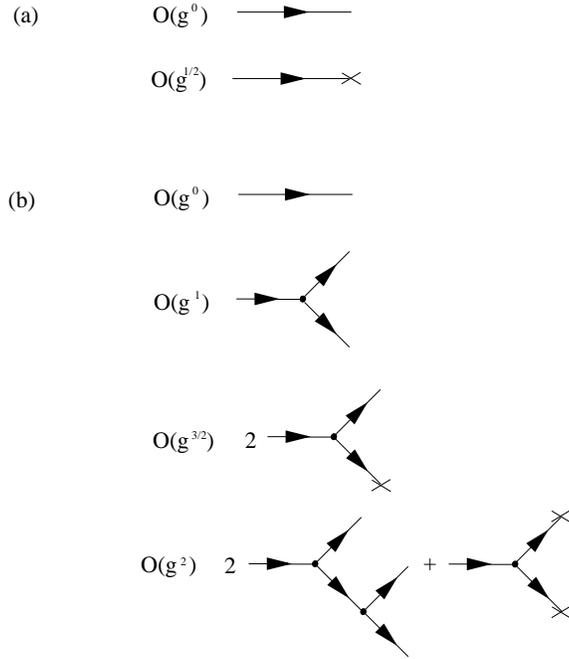,width=75mm}} 
\caption{Stochastic diagrams representing the lowest order nonzero 
terms required to determine the intracavity moment of quantum mechanics
$\langle {\hat M}(\tau) \rangle_{QM}$
for (a) $\alpha_{1}(\tau), \alpha_{1}^{\dag}(\tau), \alpha_{2}(\tau)\:
{\rm and}\:
\alpha_{2}^{\dag}(\tau)$ and (b) $\alpha_{3}(\tau),\alpha_{3}^{\dag}(\tau),
\langle\alpha_{3}(\tau)\rangle \:{\rm and}\:
\langle\alpha_{3}^{\dag}(\tau)\rangle$.}
\label{fig2}
\end{figure}

Figs~\ref{fig2} (a) and (b) show the stochastic diagrams
required to determine the moments of the form
$\langle {\cal A}_{1}(\tau){\cal A}_{2}(\tau)\Delta {\cal A}_{3}(\tau)\rangle$. 
Naively, it might be thought that the lowest order nonzero terms from  
${\cal A}_{1}(\tau),\:{\cal A}_{2}(\tau)\:{\rm and}\:\Delta\:{\cal A}_{3}(\tau)$
simply need to be multiplied together and the average of the subsequent
product determined to calculate the lowest order nonzero term
in $\langle {\cal A}_{1}(\tau)\:{\cal A}_{2}(\tau)\Delta\:{\cal A}_{3}(\tau) 
\rangle$. 
This is not always true. Sometimes,
${\cal A}_{1}(\tau),\:{\cal A}_{2}(\tau)\:{\rm and}\:\Delta
\:{\cal A}_{3}(\tau)$
are not necessarily zero and yet $\langle {\cal A}_{1}(\tau){\cal A}_{2}(\tau)
\Delta {\cal A}_{3}(\tau)\rangle$
is zero.
For example, the lowest order nonzero terms for the positive-P variables in 
$\langle\alpha_{1}(\tau)\alpha_{2}(\tau) \Delta\alpha_{3}^{+}(\tau)\rangle$
are
\begin{equation}
\alpha_{i}^{[1/2]}(\tau)= 
\int_{\tau_{i}=0}^{\tau_{i}=\tau}d\tau_{i}\exp[\gamma_{i}(\tau_{i}-\tau)]
\sqrt{g\epsilon}
\exp(-\frac{\gamma_{i}\tau_{i}}{2})\xi_{i} (\tau_{i}),
\end{equation}
where $i=1,2$ and 
\begin{eqnarray}
\Delta \alpha_{3}^{+\:[2]}(\tau)&=&\int_{\tau_{3}=0}^{\tau_{3}=\tau}d\tau_{3}
\exp[\gamma_{3}(\tau_{3}-\tau)]g\int_{\tau_{4}=0}^{\tau_{4}=\tau_{3}}d\tau_{4}
\exp[\gamma_{1}(\tau_{4}-\tau_{3})]\sqrt{g\epsilon^{*}}
\exp(-\frac{\gamma_{1}\tau_{4}}{2})\xi_{1}^{+} (\tau_{4}) \\ 
& & \int_{\tau_{5}=0}^{\tau_{5}=\tau_{3}}d\tau_{5}\exp[\gamma_{2}(\tau_{5}-
\tau_{3})]
\sqrt{g\epsilon^{*}}\exp(-\frac{\gamma_{2}\tau_{5}}{2})\xi_{2}^{+} (\tau_{5}) 
\\ \nonumber 
& & -\bigl\langle\int_{\tau_{3}=0}^{\tau_{3}=\tau}d\tau_{3}
\exp[\gamma_{3}(\tau_{3}-\tau)]g\int_{\tau_{4}=0}^{\tau_{4}=\tau_{3}}d\tau_{4}
\exp[\gamma_{1}(\tau_{4}-\tau_{3})]\sqrt{g\epsilon^{*}}
\exp(-\frac{\gamma_{1}\tau_{4}}{2})\xi_{1}^{+} (\tau_{4}) \\ \nonumber
& & \int_{\tau_{5}=0}^{\tau_{5}=\tau_{3}}d\tau_{5}\exp[\gamma_{2}(\tau_{5}-
\tau_{3})]
\sqrt{g\epsilon^{*}}\exp(-\frac{\gamma_{2}\tau_{5}}{2})\xi_{2}^{+}
(\tau_{5})\bigr\rangle.
\end{eqnarray} 
However, the average of their product is zero as 
\begin{eqnarray}
& &\bigl
\langle\alpha_{1}^{[1/2]}(\tau)\alpha_{2}^{[1/2]}(\tau)\Delta\alpha_{3}^{+\:[2]}
(\tau)\bigr
\rangle \\ \nonumber
&=&g^{3} \epsilon \epsilon^{*} \int_{\tau_{1}=0}^{\tau_{1}=\tau}
\int_{\tau_{2}=0}^{\tau_{2}=\tau}
\int_{\tau_{3}=0}^{\tau_{3}=\tau}\int_{\tau_{4}=0}^{\tau_{4}=\tau_{3}}
\int_{\tau_{5}=0}^{\tau_{5}=\tau_{3}} d\tau_{1} d\tau_{2}
d\tau_{3} d\tau_{4} d\tau_{5} \\ \nonumber
& & \exp[\gamma_{1}(\tau_{1}-\tau)]
\exp(-\frac{\gamma_{1}\tau_{1}}{2})\exp[\gamma_{2}(\tau_{2}-\tau)]
\exp(-\frac{\gamma_{2}\tau_{2}}{2})\exp[\gamma_{3}(\tau_{3}-\tau)]
\exp[\gamma_{1}(\tau_{4}-\tau_{3})]
\exp(-\frac{\gamma_{1}\tau_{4}}{2})\\ \nonumber
& &\exp[\gamma_{5}(\tau_{5}-\tau_{3})]
\exp(-\frac{\gamma_{2}\tau_{5}}{2}) 
\bigl(\langle\xi_{1}(\tau_{1})\xi_{2}(\tau_{2})\xi_{1}^{+}(\tau_{4})
\xi_{2}^{+}(\tau_{5})\rangle
-\langle\xi_{1}(\tau_{1})\xi_{2}(\tau_{2})\rangle\:
\langle\xi_{1}^{+}(\tau_{4})\xi_{2}^{+}(\tau_{5})\rangle\bigl) \nonumber
\end{eqnarray}
and
\begin{equation} \label{cancel}
\langle\xi_{1}(\tau_{1})\xi_{2}(\tau_{2})\xi_{1}^{+}(\tau_{4})\xi_{2}^{+}
(\tau_{5})\rangle
=\langle\xi_{1}(\tau_{1})\xi_{2}(\tau_{2})\rangle\:\langle\xi_{1}^{+}(\tau_{4})
\xi_{2}^{+}(\tau_{5})\rangle.
\end{equation}
Thus, the two noise terms cancel each other and the  
right hand side of Eq.~(\ref{cancel}) is zero.
Taking such a consideration into account, the  
moments of the form $\langle {\cal A}_{1}(\tau){\cal A}_{2}(\tau)
\Delta {\cal A}_{3}(\tau)\rangle$  are determined by
carefully considering the lowest order nonzero terms of their 
constituent positive-P variables 
and then finding the average of these variables' products.

Consider Fig.~(\ref{fig2})(a), which contains
the lowest order stochastic diagrams for $\alpha_{i}(\tau)$ and 
$\alpha_{i}^{+}(\tau)$, where $i=1,2$.
The first diagram in it represents the initial value terms 
$\alpha_{i}^{(0)}(\tau)$
and $\alpha_{i}^{(0)\:+}(\tau)$, which are zero and do not
contribute to any moments. The second represents 
terms containing
$\xi_{1},\xi_{1}^{+},\xi_{2}$ or $\xi_{2}^{+}$, which
are not necessarily zero and thus may contribute to moments.
Fig.~\ref{fig2}(b) contains the lowest order stochastic diagrams for 
$\alpha_{3}(\tau)$ and $\alpha_{3}^{+}(\tau)$.
In it, all terms represented by stochastic diagrams containing initial
value lines are zero
except for the $O(g^{0})$ ones. 
This is so as these terms contain either $\alpha_{i}^{(0)}(\tau)\:{\rm or}\:
\alpha_{i}^{+\:(0)}(\tau)$, 
where $i=1,2$, which are both zero. In addition, all $O(g^{0})$ terms 
represented by stochastic diagrams in Fig.~\ref{fig2}(b) are 
canceled out 
by other $O(g^{0})$ terms. This occurs because
$\Delta {\cal A}_{3}(\tau)$ appears in the moments considered.
Its two components, ${\cal A}_{3}(\tau)$ and $\langle {\cal A}_{3}(\tau)
\rangle$, 
contain the same $O(g^{0})$ term and hence their $O(g^{0})$ terms cancel each 
other. It follows that the only remaining stochastic diagram in 
Fig.~\ref{fig2}(b),
which represents the
$O(g^{2})$ term containing two noise components, denotes the lowest order 
term in $\Delta {\cal A}_{3}(\tau)$ that is not necessarily zero.

The lowest order nonzero terms determined above are now used to calculate
$\langle\alpha_{1}(\tau)\alpha_{2}(\tau) \Delta \alpha_{3}(\tau)\rangle$.
The lowest order contribution to $\Delta \alpha_{3}(\tau)$ that is not
necessarily zero $\Delta \alpha_{3\:{\rm lowest}}(\tau)$ is
\begin{eqnarray}
\Delta \alpha_{3\:{\rm lowest}}(\tau)&=&\int_{\tau_{3}=0}^{\tau_{3}=\tau}
d\tau_{3}
\exp[\gamma_{3}(\tau_{3}-\tau)]g\int_{\tau_{4}=0}^{\tau_{4}=\tau_{3}}d\tau_{4}
\exp[\gamma_{1}(\tau_{4}-\tau_{3})]\sqrt{g\epsilon}
\exp(-\frac{\gamma_{1}\tau_{4}}{2})\xi_{1}(\tau_{4}) \\ \nonumber
& & \int_{\tau_{5}=0}^{\tau_{5}=\tau_{3}}d\tau_{5}\exp[\gamma_{2}
(\tau_{5}-\tau_{3})]
\sqrt{g\epsilon}\exp(-\frac{\gamma_{2}\tau_{5}}{2})\xi_{2}(\tau_{5}) 
\\ \nonumber 
& & -\bigl\langle\int_{\tau_{3}=0}^{\tau_{3}=\tau}d\tau_{3}
\exp[\gamma_{3}(\tau_{3}-\tau)]g\int_{\tau_{4}=0}^{\tau_{4}=\tau_{3}}
d\tau_{4}\exp[\gamma_{1}(\tau_{4}-\tau_{3})]\sqrt{g\epsilon}
\exp(-\frac{\gamma_{1}\tau_{4}}{2})\xi_{1}(\tau_{4}) \\ \nonumber
& & \int_{\tau_{5}=0}^{\tau_{5}=\tau_{3}}d\tau_{5}\exp[\gamma_{2}
(\tau_{5}-\tau_{3})]
\sqrt{g\epsilon}\exp(-\frac{\gamma_{2}\tau_{5}}{2})\xi_{2}
(\tau_{5})\bigr\rangle. 
\end{eqnarray}
When $\gamma_{1}=\gamma_{2}=\gamma_{3}=\gamma$, the 
average of the product of $\Delta \alpha_{3\:{\rm lowest}}(\tau)$ and the 
lowest order nonzero 
terms in $\alpha_{1}(\tau)$ and $\alpha_{2}(\tau)$ is
approximately equal to
$\langle \alpha_{1}(\tau) \alpha_{2}(\tau)\Delta \alpha_{3}
(\tau)\rangle$ when $g \ll 1$ and thus
\begin{eqnarray}
\langle \alpha_{1}(\tau) \alpha_{2}(\tau)\Delta \alpha_{3}
(\tau)\rangle &\simeq&
-\frac{\epsilon^{2} g^{3} \exp[-3\gamma \tau]}{\gamma^{2}} 
\times \bigl[\frac{\exp[\gamma \tau]}{\gamma}-2\tau-
\frac{\exp[-\gamma \tau]}{\gamma}\bigr].
\label{leading_2}
\end{eqnarray}
As daggered positive-P variables are complex conjugate to undaggered ones
on average, $\langle \alpha_{1}^{+}(\tau) \alpha_{2}^{+}
(\tau)\Delta \alpha_{3}^{+}(\tau)\rangle=\langle \alpha_{1}(\tau)
\alpha_{2}(\tau)\Delta \alpha_{3}(\tau)\rangle^{*}$.
The other six moments of the form $\langle {\cal A}_{1}(\tau){\cal A}_{2}(\tau)
\Delta {\cal A}_{3}(\tau)\rangle$ are zero to $O(g^{3})$. 
As does $\langle \alpha_{1}(\tau) \alpha_{2}(\tau)
\Delta\alpha_{3}^{+}(\tau)\rangle$,
they all have two $O(g^{3})$ terms that cancel each other. To explain such
behaviour in general, the following argument is given.
These other six moments can be rewritten as
\begin{equation} \label{star}
\langle {\cal A}_{1}(\tau){\cal A}_{2}(\tau){\cal A}_{3}(\tau) \rangle - 
\langle {\cal A}_{1}(\tau) {\cal A}_{2}(\tau)\rangle\langle
{\cal A}_{3}(\tau)\rangle, 
\end{equation}
where it is understood that the moments in which 
${\cal A}_{1}(\tau)=\alpha_{1}(\tau), {\cal A}_{2}(\tau)=\alpha_{2}(\tau), 
{\cal A}_{3}(\tau)=\alpha_{3}(\tau)$ and
${\cal A}_{1}(\tau)=\alpha_{1}^{+}(\tau), {\cal
A}_{2}(\tau)=\alpha_{2}^{+}(\tau),
{\cal A}_{3}(\tau)=\alpha_{3}^{+}(\tau)$
are excluded. All $O(g^{3})$ terms in the six moments of the form 
$\langle {\cal A}_{1}(\tau){\cal A}_{2}(\tau){\cal A}_{3}(\tau) \rangle$ 
under consideration contain 
noises in one of the following three forms, 
$\bigl\langle \xi_{1}(\tau_{a})\xi_{2}(\tau_{b})\xi_{1}^{\dag}(\tau_{c})
\xi_{2}^{\dag}(\tau_{d}) \bigr \rangle$,
$\bigl\langle \xi_{i}(\tau_{a})\xi_{3-i}^{\dag}(\tau_{b})\xi_{i}(\tau_{c})
\xi_{3-i}(\tau_{d})\bigr\rangle$ and 
$\bigl\langle \xi_{i}^{\dag}(\tau_{a})\xi_{3-i}(\tau_{b})\xi_{i}^{\dag}
(\tau_{c})
\xi_{3-i}^{\dag}(\tau_{d})\bigr\rangle$,
where $i=1,2$ and time arguments are dummy variables.
All $O(g^{3})$ terms in the six moments of the form 
$\langle {\cal A}_{1}(\tau){\cal A}_{2}(\tau)\rangle \langle{\cal A}_{3}(\tau) 
\rangle$ under consideration
contain the same noises as their corresponding
$\langle {\cal A}_{1}(\tau){\cal A}_{2}(\tau){\cal A}_{3}(\tau) \rangle$
term. However, in these terms of the form 
$\langle {\cal A}_{1}(\tau){\cal A}_{2}(\tau)\rangle \langle{\cal A}_{3}(\tau) 
\rangle$
four noise averages from 
corresponding terms of the form 
$\langle {\cal A}_{1}(\tau){\cal A}_{2}(\tau){\cal A}_{3}(\tau) \rangle$
are split into the product of two averages of two noises.
For example,
$\langle \alpha_{1}(\tau)\alpha_{2}(\tau)\alpha_{3}^{+}(\tau)\rangle$
contains noises in the form 
$\langle \xi_{1}(\tau_{a})\xi_{2}(\tau_{b})\xi_{1}^{+}(\tau_{c})
\xi_{2}^{+}(\tau_{d})\rangle$
whilst $\langle \alpha_{1}(\tau)\alpha_{2}(\tau)\rangle 
\langle \alpha_{3}^{+}(\tau)\rangle$ contains them in the form
$\langle \xi_{1}(\tau_{a})\xi_{2}(\tau_{b})\rangle 
\langle \xi_{1}^{+}(\tau_{c}) \xi_{2}^{+}(\tau_{d})\rangle$.
Using the formula
\begin{eqnarray}
\bigl\langle \xi_{1}(\tau_{a})\xi_{2}(\tau_{b})\xi_{3}(\tau_{c})\xi_{4}
(\tau_{d})\bigr\rangle 
&=&\:\:\:\bigl\langle \xi_{1}(\tau_{a})\xi_{2}(\tau_{b})\bigr\rangle
\bigl\langle
\xi_{3}(\tau_{c})\xi_{4}(\tau_{d})\bigr\rangle \\ \nonumber
& &+ \bigl\langle \xi_{1}(\tau_{a})\xi_{3}(\tau_{c})\bigr\rangle\:\bigl\langle
\xi_{2}(\tau_{b})\xi_{4}(\tau_{d})\bigr\rangle \\ \nonumber
& &+ \bigl\langle \xi_{1}(\tau_{a})\xi_{4}(\tau_{d})\bigr\rangle\:\bigl\langle
\xi_{2}(\tau_{b})\xi_{3}(\tau_{c})\bigr\rangle   
\end{eqnarray}
it can be shown that noise expressions in moments of the form 
$\langle {\cal A}_{1}(\tau){\cal A}_{2}(\tau) {\cal A}_{3}(\tau) \rangle$
under consideration
factorize. In particular, they reduce to the noise expression
in the six corresponding term of the form
$\langle {\cal A}_{1}(\tau){\cal A}_{2}(\tau)\rangle \langle{\cal A}_{3}(\tau) 
\rangle$. It follows that cancellation occurs between the $O(g^{3})$ terms in 
corresponding moments of the form 
$\langle {\cal A}_{1}(\tau){\cal A}_{2}(\tau) {\cal A}_{3}(\tau) \rangle$ and 
$\langle {\cal A}_{1}(\tau){\cal A}_{2}(\tau)\rangle \langle{\cal A}_{3}(\tau) 
\rangle$ under consideration
as the two terms are identical.
Consequently, all six moments under consideration are $O(g^{4})$. 
They are also typically much smaller than the two $O(g^{3})$ moments,
$\langle \alpha_{1}(\tau) \alpha_{2}(\tau) \Delta\alpha_{3}(\tau)\rangle$ 
and $\langle \alpha_{1}^{+}(\tau) \alpha_{2}^{+}(\tau) 
\Delta \alpha_{3}^{+}(\tau) \rangle$,
as $g \ll 1$ for realistic systems.
To be precise, as all moments are complex quantities, 
the magnitudes of 
$\langle \alpha_{1}(\tau) \alpha_{2}(\tau) \Delta\alpha_{3}(\tau)\rangle$ 
and $\langle \alpha_{1}^{+}(\tau) \alpha_{2}^{+}(\tau) 
\Delta \alpha_{3}^{+}(\tau) \rangle$ are much larger than the magnitudes of
the other six moments.

\begin{figure}
\center{\epsfig{figure=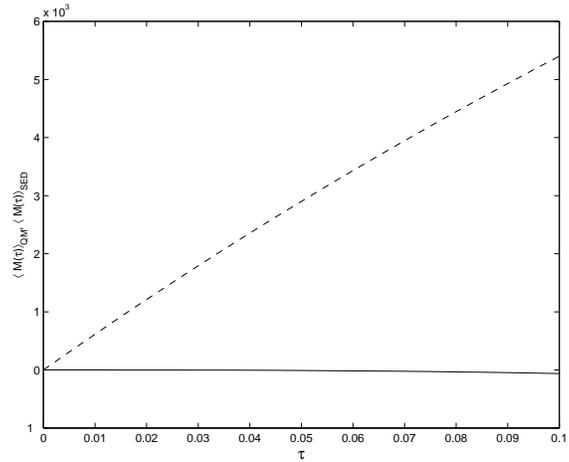,width=75mm}} 
\caption{Analytic results for $\langle {\hat M}(\tau) \rangle_{QM}$
(solid line) and $\langle M(\tau) \rangle_{SED}$ (dotted line) 
versus scaled time $\tau$
for N=1, g=1, $\gamma=1$ and $\cos\Theta=\cos\Phi=1$.}
\label{fig3}
\end{figure}

The above results are now more closely related to experiments by considering
quadrature phase amplitudes $X_{i,\theta_{i}}(\tau)$.
In particular, calculations are performed to determine
the in principle experimentally observable third order quadrature phase 
amplitude moment
$\langle M(\tau) \rangle$, where $\langle M(\tau) \rangle=
\langle \Delta X_{1,\theta_{1}}(\tau) 
\Delta X_{2,\theta_{2}}(\tau)
\Delta X_{3,\theta_{3}}(\tau)\rangle$, according to quantum mechanics and SED.
In quantum mechanics, 
quadrature phase amplitudes are expressed in terms of creation and annihilation operators by the
equation 
\begin{equation} \label{quad_eqn}
{\hat X}_{i,\theta_{i}}=\frac{{\hat a}_{i}\exp(-i\theta_{i})+
{\hat a}_{i}^{\dag}\exp(i\theta_{i})}{2}.
\end{equation}
Using Eq.~(\ref{quad_eqn}) and operator-positive-P variable correspondences
$\langle {\hat M}(\tau) \rangle_{QM}$, the value of
$\langle M(\tau) \rangle$ for quantum mechanics can be expressed as 
\begin{equation} \label{M_qm}
\langle {\hat M}(\tau)\rangle_{QM}
=\frac{1}{8}\langle\prod^{3}_{i=1}\Delta \alpha_{i}(\tau)e^{-i\theta_{i}}+
\Delta
\alpha^{+}_{i}(\tau)e^{i\theta_{i}}\rangle.
\label{triple}
\end{equation}
Upon expanding the right hand side of Eq.~(\ref{M_qm}),
the two lowest order terms in g,  
$\langle \Delta \alpha_{1}(\tau) \Delta \alpha_{2}(\tau) 
\Delta\alpha_{3}(\tau)\rangle$ 
and $\langle \Delta \alpha_{1}^{+}(\tau) \Delta \alpha_{2}^{+}(\tau) 
\Delta \alpha_{3}^{+}(\tau) \rangle$, usually dominate. When they do
\begin{equation} \label{quad}
\langle {\hat M}(\tau) \rangle_{QM}
\simeq\frac{1}{4}[\cos\Theta{\rm Re\langle \alpha_{1}(\tau)
\alpha_{2}(\tau)\Delta \alpha_{3}(\tau)\rangle}
-\sin\Theta{\rm Im\langle \alpha_{1}(\tau) \alpha_{2}(\tau)\Delta \alpha_{3}
(\tau)\rangle}],
\end{equation}
where $\Theta=\theta_{1}+\theta_{2}+\theta_{3}$.
However, when 
${\rm cos}\Theta=0$ 
and ${\rm Im}(\langle \alpha_{1}(\tau) \alpha_{2}(\tau)\Delta\alpha_{3}
(\tau)\rangle)
=O(\tau^{4})$ or when
${\rm sin}\Theta=0$ and 
${\rm Re}(\langle \alpha_{1}(\tau)
\alpha_{2}(\tau)\Delta\alpha_{3}(\tau)\rangle)=O(\tau^{4})$
Eq.~(\ref{quad}) is not necessarily true.
Such situations can be avoided 
though because $\Theta$ and $\epsilon$ are controllable parameters.
They are ignored in present considerations. When $\epsilon$ is real,
the $O(g^{3})$ term in
$\langle\alpha_{1}(\tau) \alpha_{2}(\tau)\Delta \alpha_{3}(\tau)\rangle$
is also real and so 
\begin{equation} \label{quantum_quadrature}
\langle {\hat M}(\tau) \rangle_{QM} \simeq\frac{1}{4} \cos\Theta \:\langle 
\alpha_{1}(\tau)
\alpha_{2}(\tau)\Delta \alpha_{3}(\tau) \rangle \simeq
-\frac{\epsilon^{2}g^{3}\cos\Theta \exp(-3 \gamma \tau)}{4 \gamma^{2}}
\bigl[ \frac{\exp(\gamma \tau)}{\gamma}- 2\tau - \frac{\exp(-\gamma \tau)}
{\gamma}\bigr].
\end{equation}
Thus, Eq.~(\ref{quantum_quadrature})
shows that $\langle {\hat M}(\tau)\rangle_{QM}$ is cubic in g,
within the domain considered, as shown in Fig.~\ref{fig3}.

\section{comparison of quantum mechanics and stochastic electrodynamics} 
\label{sed_section}

This section compares the predictions of quantum mechanics and SED for 
the intracavity moment $\langle M(\tau) \rangle$. SED is a semiclassical theory which adds
Gaussian white noise to classical electrodynamics. It describes electromagnetic
field modes by complex field amplitudes $\beta$. For the
nondegenerate parametric oscillator, the set of such amplitudes
$\{\beta_{1},\beta_{2},\beta_{3}\}$ evolves via the equations  
\begin{eqnarray} \label{sed}
\frac{\partial \beta_{1}}{\partial \tau} &=& -\gamma_{1}
\beta_{1}+g\beta_{2}^{*}\beta_{3}+\sqrt{\gamma_{1}} \xi_{1} \\ \nonumber
\frac{\partial \beta_{2}}{\partial \tau} &=& -\gamma_{2} \beta_{2}
+g\beta_{1}^{*}\beta_{3}+\sqrt{\gamma_{2}} \xi_{2} \\ \nonumber
\frac{\partial \beta_{3}}{\partial \tau} &=& -\gamma_{3} \beta_{3}
-g\beta_{1}\beta_{2}+\sqrt{\gamma_{3}} \xi_{3},
\end{eqnarray}
where the same time variable as in the quantum case is used and the $\xi's$ are independent complex Gaussian 
white noises with the following correlations
\begin{equation}
\langle\xi_{i}(\tau_{1})\:
\xi_{j}^{*}(\tau_{2})\rangle=\delta_{ij}\delta(\tau_{1}-\tau_{2}),
\end{equation}
where $i,j=1,2,3$.
The field amplitudes $\beta_{1},\beta_{2}\:{\rm and}\:\beta_{3}$ initially 
have Gaussian fluctuations in their real and imaginary parts
of variance 1/4. The only nonzero correlations present in these  
fluctuations are thus 
\begin{equation}
\langle\Delta\beta_{i}(0)\Delta\beta_{i}^{*}(0)\rangle=\frac{1}{2},
\end{equation}
where $i=1,2,3$. Initial conditions are
$\langle\beta_{1}(0)\rangle=\langle\beta_{2}(0)\rangle=0 \:{\rm and}\:
\langle\beta_{3}(0)\rangle=\epsilon$.

The SED prediction for the intracavity moment
$\langle M(\tau) \rangle$ is $\langle M(\tau)\rangle_{SED}$, 
which is given by the equation 
\begin{equation}
\langle M(\tau) \rangle_{SED}=
\langle \Delta X_{1,\theta_{1}}(\tau) \Delta X_{2,\theta_{2}}(\tau) 
\Delta X_{3\:\theta_{3}}(\tau) \rangle,
\end{equation}
where $X_{i,\theta_{i}}(\tau) = \bigl(\beta_{i}(\tau) e^{-i\theta_{i}} + 
\beta_{i}^{*}
(\tau) e^{i\theta_{i}}\bigr)/2$.
It is calculated using 
a similar iterative method to the one in Section \ref{section_two}, 
except that noise terms are 
now treated exactly instead of iteratively. Zeroth order approximations for 
this iterative method are thus
\begin{eqnarray}
\beta_{i}^{(0)}(\tau)&=&\beta_{i}(0)\exp(-\gamma_{i}\tau)+
\int_{\tau_{i=0}}^{\tau_{i}=\tau}
d\tau_{i} \exp[\gamma_{i}(\tau_{i}-\tau)]\sqrt{\gamma_{i}}\xi_{i} 
\\ \nonumber
\beta_{3}^{(0)}(\tau)&=&\beta_{3}(0)\exp(-\gamma_{3}\tau),
\end{eqnarray}
where $i=1,2$. Higher order $(m+1)^{th}$ order approximations are 
\begin{eqnarray}
\beta_{i}^{(m+1)}(\tau)&=&\beta_{i}^{(0)}(\tau)+
\int^{\tau_{i}=\tau}_{\tau_{i}=0}
d\tau_{i} \exp[\gamma_{i}(\tau_{i}-\tau)]g \beta_{3-i}^{*\:(m)}(\tau_{i})
\beta_{3}^{(m)}(\tau_{i})\\ \nonumber
\beta_{3}^{(m+1)}(\tau)&=&\beta_{3}^{(0)}(\tau)-
\int^{\tau_{3}=\tau}_{\tau_{3}=0}
d\tau_{3} \exp[\gamma_{3}(\tau_{3}-\tau)]g \beta_{1}^{(m)}(\tau_{3})
\beta_{2}^{(m)}(\tau_{3}),
\end{eqnarray}
where $i=1,2$.

\begin{figure}
\center{\epsfig{figure=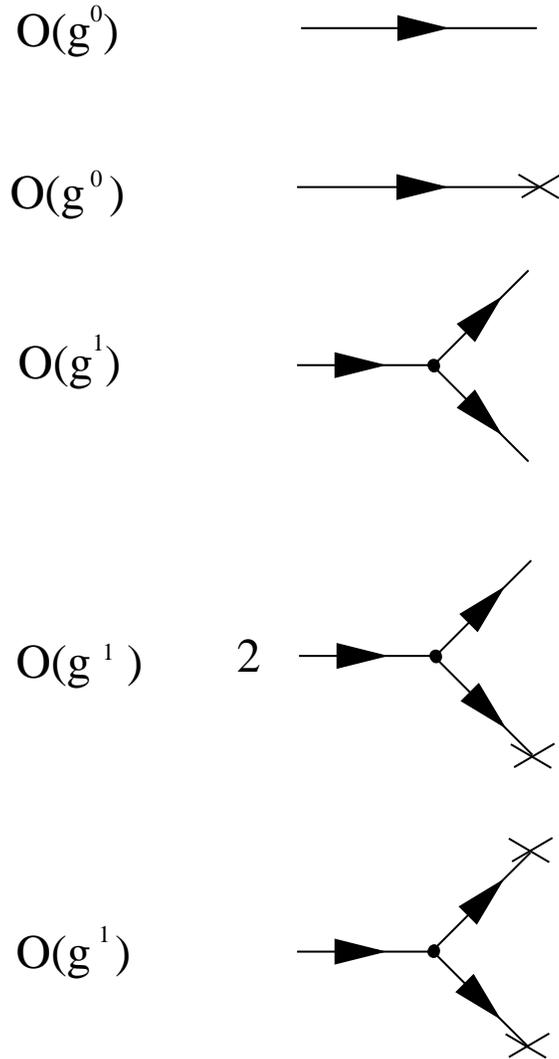,width=75mm}} 
\caption{Stochastic diagrams representing the lowest order nonzero
terms required to determine the intracavity moment of SED 
$\langle M(\tau) \rangle_{SED}$
for $\beta_{i}$, where $i=1,2,3$.}
\label{fig4}
\end{figure}

The lowest order nonzero term in g of $\langle M(\tau) \rangle_{SED}$
is now found using the same method as for the lowest order nonzero term of
$\langle {\hat M}(\tau) \rangle_{QM}$. Consider the moments of the form 
$\langle\Delta {\cal B}_{1}(\tau)\Delta {\cal B}_{2}(\tau) 
\Delta {\cal B}_{3}(\tau)\rangle$, 
where ${\cal B}_{n}(\tau)$ is either $\beta_{n}(\tau)$ or
$\beta_{n}^{*}(\tau)$. The stochastic diagrams required to determine the order 
of the lowest order nonzero
terms of these moments are shown in Fig.~\ref{fig4}. Note
that noise terms are now $O(g^{0})$, instead of $O(g^{\frac{1}{2}})$ as for
quantum mechanics. Using the stochastic diagrams in Fig.~\ref{fig4} it is 
found that, when $\gamma_{1}=\gamma_{2}=\gamma_{3}=\gamma$ and 
$g,\tau_{f} \ll 1$,
\begin{equation}
\langle\Delta\beta_{1}(\tau)\Delta \beta_{2}(\tau) \Delta 
\beta_{3}^{*}(\tau)\rangle \simeq \frac{g}{12\gamma}[1-\exp(-3 \gamma \tau)]
\simeq \langle\Delta\beta_{1}^{*}(\tau)\Delta \beta_{2}^{*}(\tau) \Delta 
\beta_{3}(\tau)\rangle^{*}.
\end{equation}
The other six moments of the form $\langle\Delta {\cal B}_{1}\Delta 
{\cal B}_{2}
\Delta {\cal B}_{3} \rangle$ are all $O(g^{2})$.
Thus, $\langle\Delta\beta_{1}(\tau)\Delta\beta_{2}(\tau)
\Delta\beta_{3}^{*}(\tau)\rangle$
and $\langle\Delta\beta_{1}^{*}(\tau)\Delta\beta_{2}^{*}(\tau)
\Delta\beta_{3}(\tau)\rangle$ dominate these other six moments 
when $g \ll 1$
and hence 
\begin{equation} \label{sed_result}
\langle M (\tau)\rangle_{SED} \simeq \frac{1}{4}
\cos\Phi\langle\Delta\beta_{1}(\tau)\Delta\beta_{2}(\tau)
\Delta\beta_{3}^{*}(\tau)\rangle
\simeq \frac{g \cos \Phi}{4\gamma}[1-\exp(1 - \gamma \tau)], 
\end{equation}
where $\Phi=\theta_{1}+\theta_{2}-\theta_{3}$, 
when $\Phi\neq 0$. Eq.~(\ref{sed_result}) 
shows that $\langle M(\tau) \rangle_{SED}$ is linear in g,
as shown in Fig.~\ref{fig3}. This is in contrast 
to the cubic behaviour of $\langle {\hat M} (\tau) \rangle_{QM}$.
Thus, quantum mechanics and SED predict greatly different values for 
$\langle M(\tau) \rangle$ when $g \ll 1$.

\begin{figure}
\center{\epsfig{figure=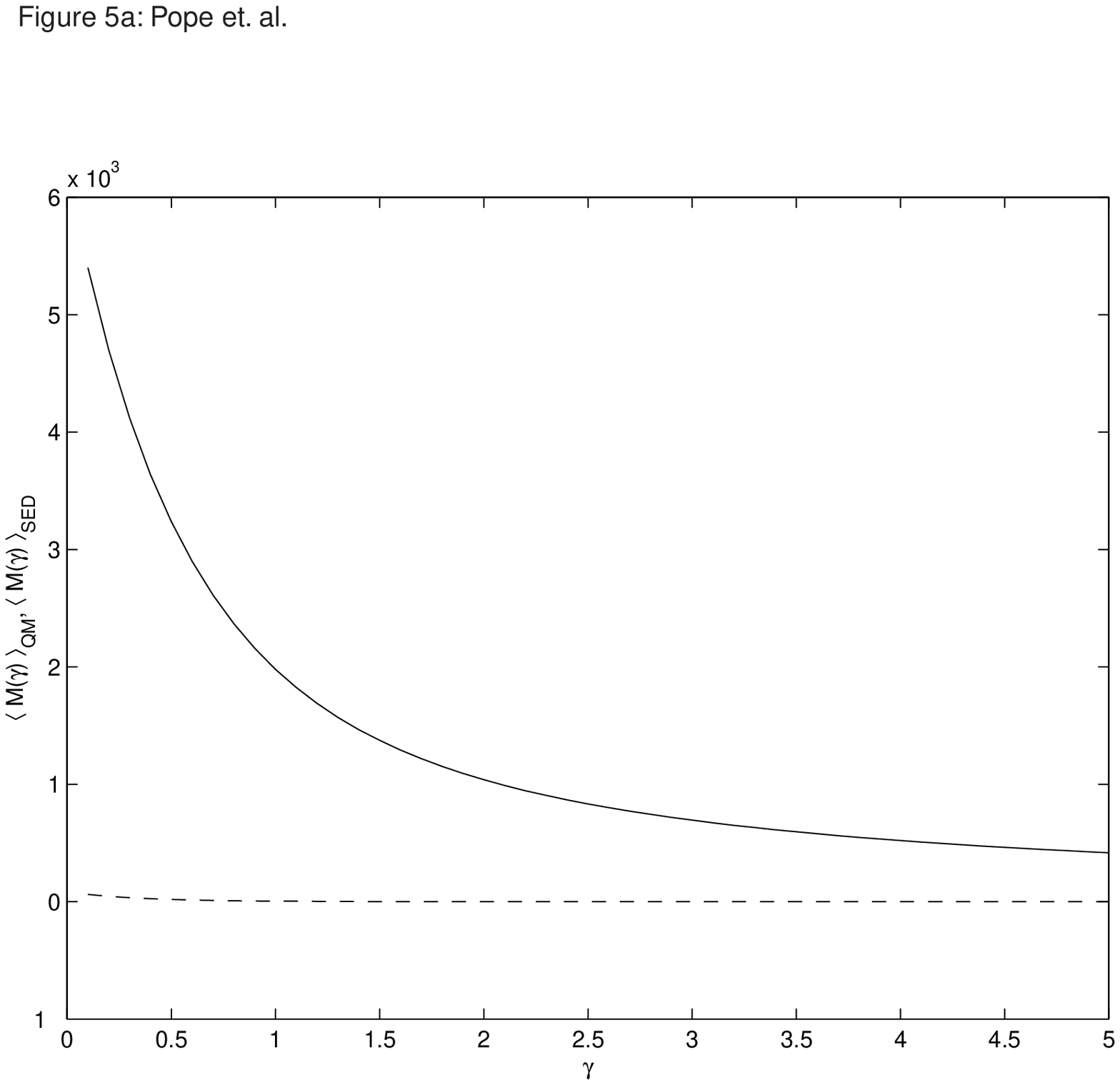,width=75mm}}
\center{\epsfig{figure=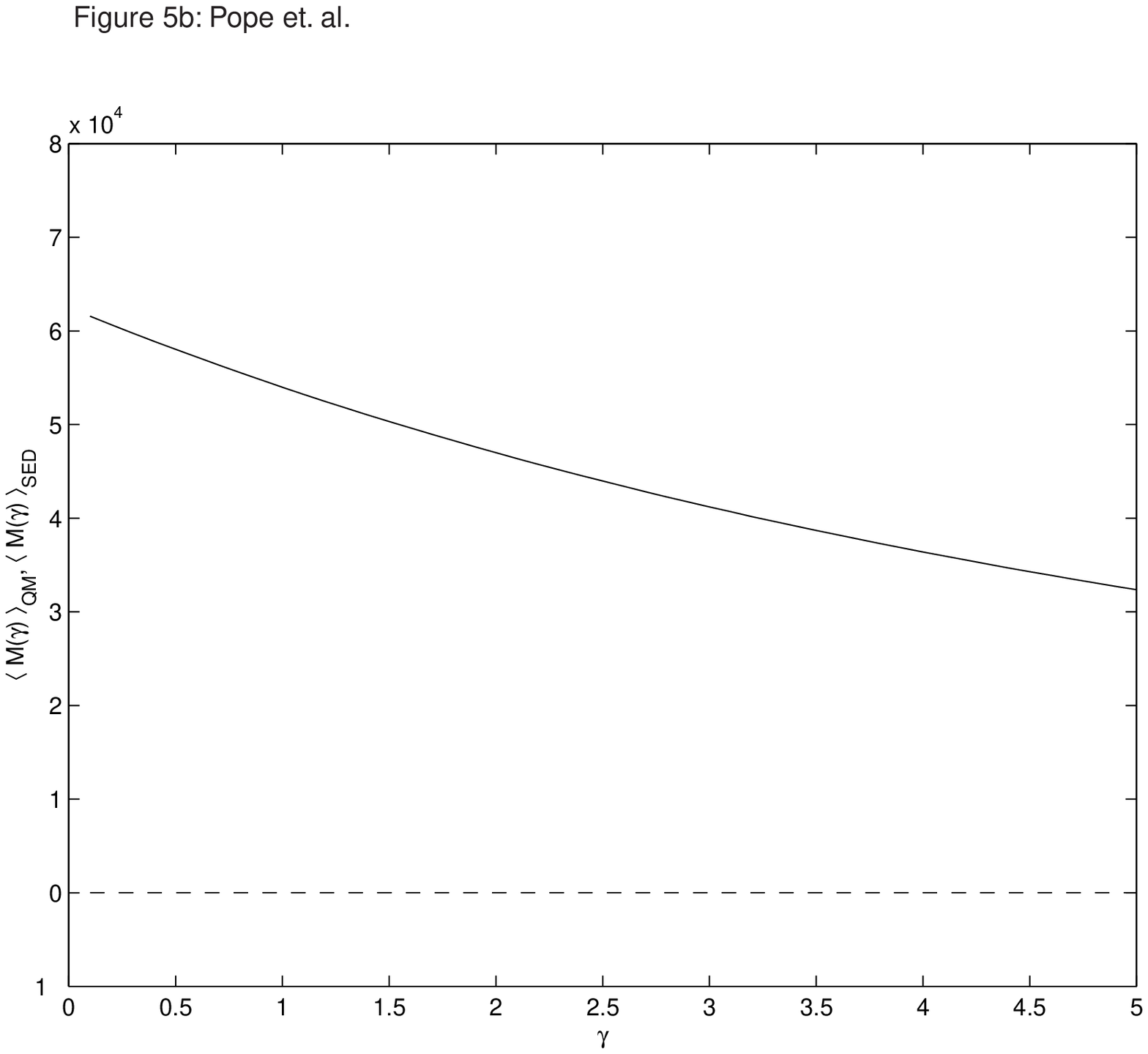,width=75mm}}
\caption{Results for $\langle {\hat M}(\gamma) \rangle_{QM}$ (dotted lines)
and $\langle M(\gamma) \rangle_{SED}$ (solid lines)
as a function of the damping constant $\gamma$ for $g=0.1, N=1$,
$\cos \Theta = \cos \Phi=0$ and (a) $\tau=1$, (b) $\tau=0.1$.}
\label{new_fig4}
\end{figure}

Consideration now is given to the effect of damping strength on the size of 
the difference between $\langle M(\tau) \rangle_{SED}$ and
$\langle {\hat M}(\tau) \rangle_{QM}$. Fig.~\ref{new_fig4} (a) shows 
$\langle {\hat M}\rangle_{QM}$ and $\langle M \rangle_{SED}$
as functions of $\gamma$ for $g=0.1,\:\tau=1$ and $\epsilon=1$. 
It indicates that the difference between them is somewhat sensitive
to $\gamma$, decreasing exponentially with increasing $\gamma$ and
quickly approaching zero. However, for the shorter time $\tau=0.1$, 
Fig.~\ref{new_fig4}(b) shows that this difference
is not as sensitive to damping. It,
approximately, only decreases linearly with increasing $\gamma$.

SED and the positive-P representation treat fluctuations very differently, 
as is evident by
comparing noise terms in Eqs~(\ref{positive_p}) and (\ref{sed}). This
difference in treatment underlies the differences between the two theories' 
results. Firstly,
noise terms in the positive-P representation are nonlinear
and are scaled by either $\sqrt{g\alpha_{3}}$ or $\sqrt{g\alpha_{3}^{+}}$, 
whilst
those in SED are linear and are scaled by $\sqrt{\gamma_{i}}$. Secondly, noise 
terms possess different correlations in the two cases. 
Thirdly, in quantum mechanics no energy fluctuations occur in the vacuum state, whilst in SED $\{ \beta_{i} \}$
fluctuates, as does the total energy. Assuming quantum mechanics is true,
in SED fluctuations in the vacuum 
lead to an overestimate of $\langle M(\tau)\rangle$ for small g.

\section{numerical results} \label{disn}

The analytic results for $\langle M(\tau) \rangle_{SED}$ and $\langle
{\hat M}(\tau)\rangle_{QM}$ 
in Sections \ref{stoch_section} and \ref{sed_section}
only include lowest order nonzero terms. This leaves the sums of all higher 
order
terms as neglected and these may be significant. For this reason,
the validity of the 
analytic approximations are checked by comparison with highly accurate 
numerical simulation results.

Numerical simulation methods for stochastic differential equations (SDE's) are 
both somewhat complex and not widely known. Thus, explanations are given for 
the
numerical technique used to solve the SDE's in Eqs~(\ref{positive_p}) and
(\ref{sed}). Normal ODE techniques such as the Runge-Kutta method cannot be 
used to solve SDE's as they contain discontinuous source terms. 
Instead, a semi-implicit numerical method 
\cite{mortimer} is employed. Only its application to Eq.~(\ref{positive_p}) 
is explained as its application Eq.~(\ref{sed}) is similar.
Each of the equations in Eq.~(\ref{positive_p}) can be rewritten as 
\begin{equation} \label{num_de}
\frac{\partial x_{i}}{\partial \tau}=A_{i}({\bf x})+\sum_{j} B_{ij}
({\bf x})\zeta_{j}(\tau),
\end{equation}
where $x_{i}$ is either $\alpha_{i}$ or $\alpha_{i}^{+}$, 
for $i=1,2,3$,
${\bf x}$ is a vector
whose components are $\{\alpha_{i},\alpha_{i}^{+}\}$, $A_{i}$ is the
function of ${\bf x}$ formed by the damping and nonlinear terms in the 
evolution
equation for $x_{i}$ and $b_{ij}$ is a matrix whose elements are 
coefficients of the 
noise terms $\{\zeta_{j}\}$ where $\zeta_{j}$ is either $\xi_{j}$ or
$\xi_{j}^{\dag}$,
for $j=1,2$.
The semi-implicit method used determines  
${\bar{{\bf x}}}^{(n)}$, an approximation to ${\bf x}$ at the midpoint of the 
interval $(\tau_{n},\tau_{n+1})$. This approximation is found 
using iteration such that the $p^{th}$ order approximation
to a component of ${\bar{{\bf x}}}^{(n)}$ $\bar{x_{i}}^{(n)\:[p]}$
is given by the equation 
\begin{equation}
\bar{x}_{i}^{(n)\:[p]}=x_{i}^{(n)}+\frac{1}{2}
\bigl[\Delta\tau A_{i}({\bf \bar{x}}^{(n)\:[p-1]})+\sum_{j}B_{ij}
({\bf \bar{x}}^{(n)\:[p-1]})\Delta W_{j}(\bar{\tau}_{n})\bigr],
\end{equation}
where $x_{i}^{(n)}$ is the value of $x_{i}$ at time $\tau_{n}$
, $\Delta\tau=\tau_{n-1}-\tau_{n}$, $\Delta W_{j}(\bar{\tau}_{n})=
\zeta_{j}^{(n)}
(\bar{\tau}_{n})\Delta\tau$
and $\bar{\tau}_{n}$ is the midpoint of the interval $(\tau_{n-1},\tau_{n})$.
The zeroth order approximation to $\bar{x}_{i}^{(n)}$
is given by the equation $\bar{x}_{i}^{(n)\:[0]}=x_{i}^{(n)}$.
The approximation to ${\bf \bar{x}}^{(n)}$ calculated
is then used to generate $\Delta x_{i}^{(n)}$, an approximation to  
the change in $x_{i}$ over the interval $(\tau_{n},\tau_{n+1})$. 
This is done by solving the equation
\begin{equation} \label{iter}
\Delta x_{i}^{(n)}=A_{i}({{\bf \bar{x}}}^{(n)})\Delta 
\tau_{n}+\sum_{j}B_{ij}({{\bf \bar{x}}}^{(n)})
\Delta W_{j}(\bar{\tau}_{n}).
\end{equation}  
Repeated use of Eq.~(\ref{iter}) determines $x_{i}^{(n)}$ for successively later
and later times and thus solves Eq.~(\ref{num_de}). 
Two of the most important parameters used in the numerical
simulations are the step size and the number of stochastic paths that are
averaged over. 
The former is always 0.0025 and the latter is $O(10^{6})$ for most simulations.
However, large sampling errors necessitated averaging over $O(10^{7})$ paths 
for $g=0.1$ SED simulations.

\begin{figure}
\center{\epsfig{figure=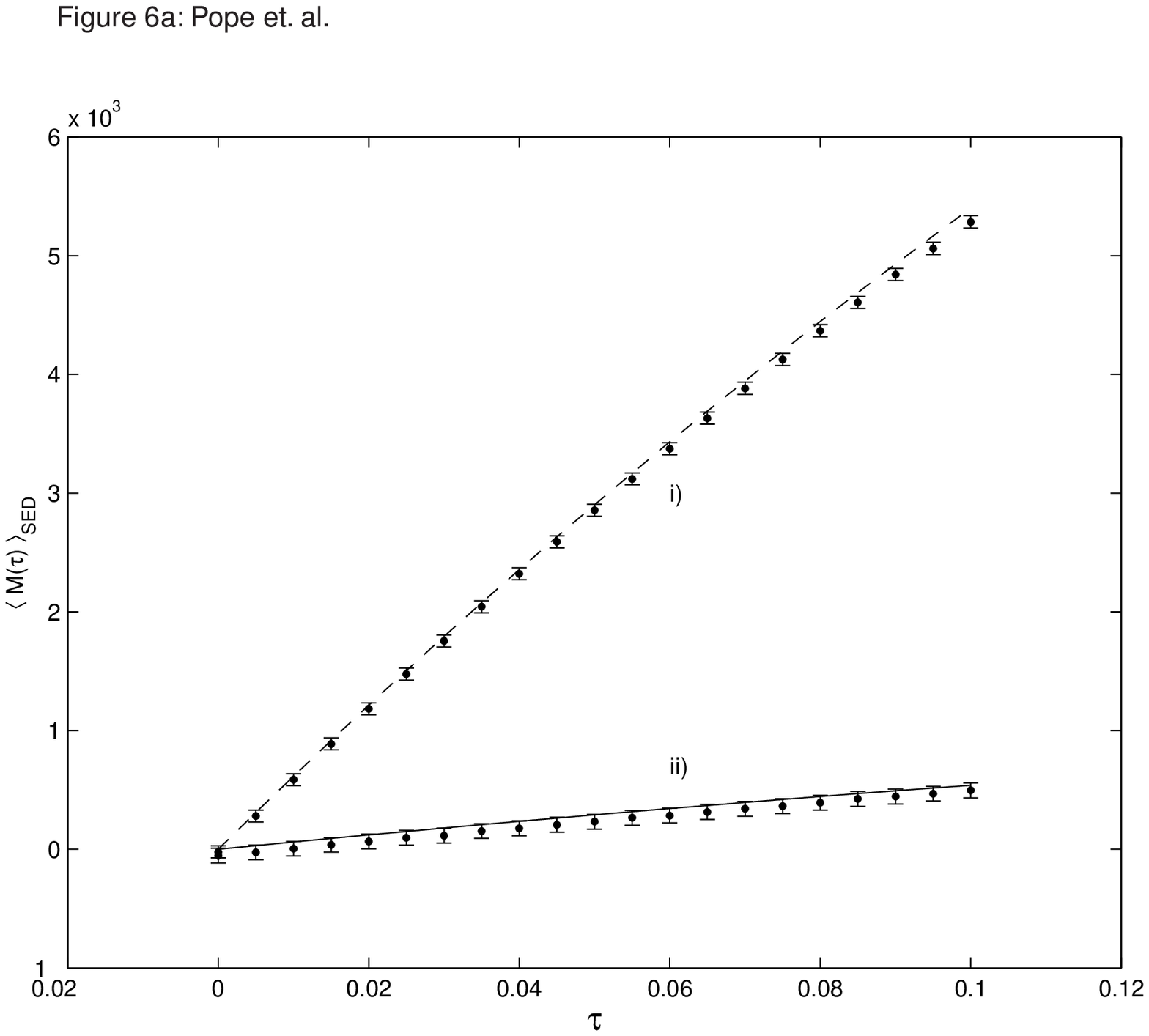,width=75mm}}
\center{\epsfig{figure=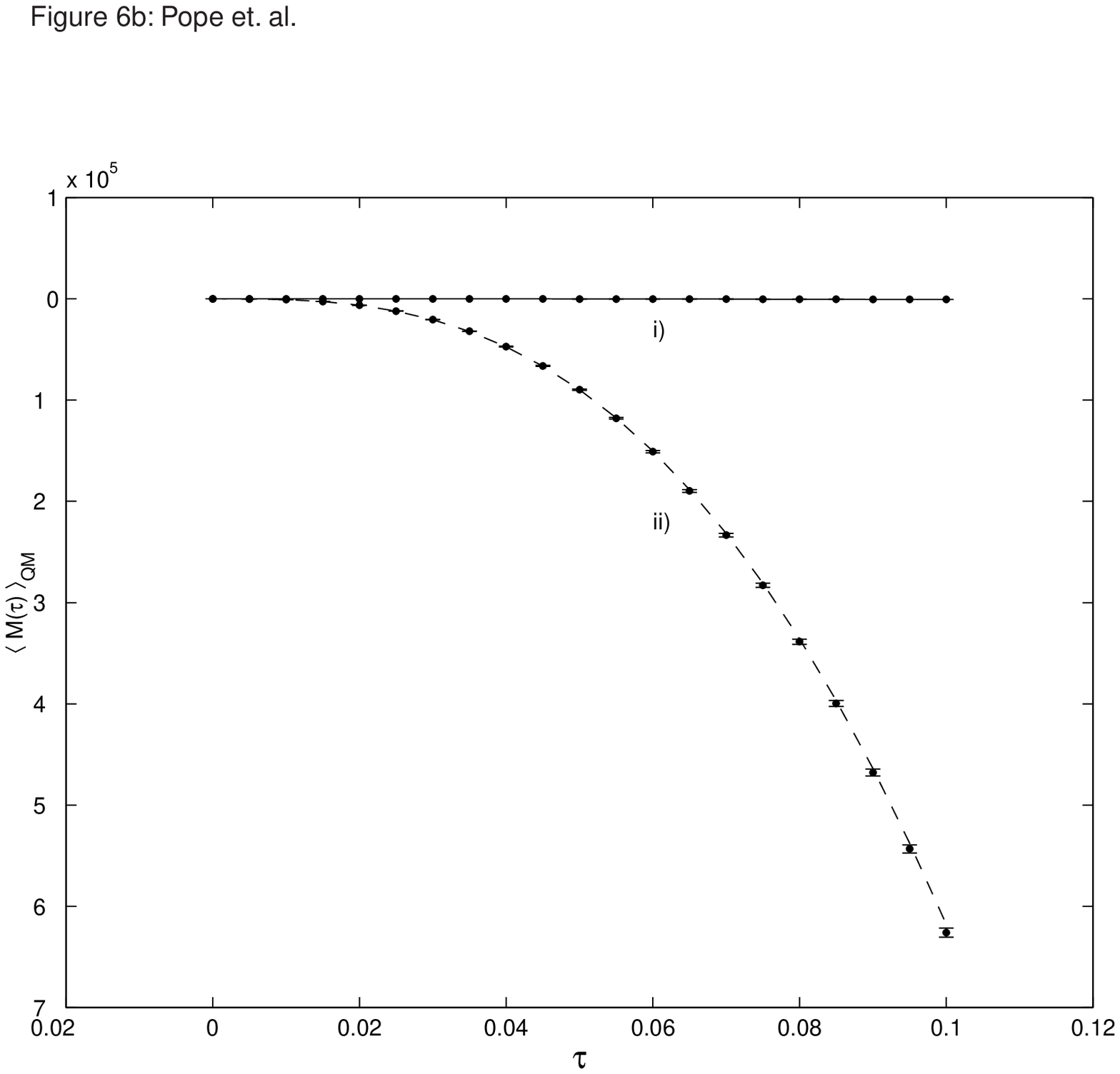,width=75mm}}
\caption{Numerical and analytic results for $\langle {\hat M}(\tau) 
\rangle_{QM}$ 
and $\langle M(\tau) \rangle_{SED}$ for $N=1$, $\gamma=1$ and 
$\cos\Theta=\cos\Phi=1$.
(a) Analytic results for $\langle M(\tau) \rangle_{SED}$ for $g=0.1$ 
(solid line)
and $g=1$ (dotted line), and numerical results for 
$\langle M(\tau) \rangle_{SED}$ represented by dots with associated error bars  
for i) $g=1$ and ii) $g=0.1$. (b) Analytic results for 
$\langle {\hat M}(\tau) \rangle_{QM}$ for $g=0.1$ (solid line)
and $g=1$ (dotted line), and numerical results for 
$\langle {\hat M}(\tau) \rangle_{QM}$ represented by dots with associated 
error bars 
for i) $g=1$ and ii) $g=0.1$.} 
\label{fig5}
\end{figure}

\begin{figure}
\center{\epsfig{figure=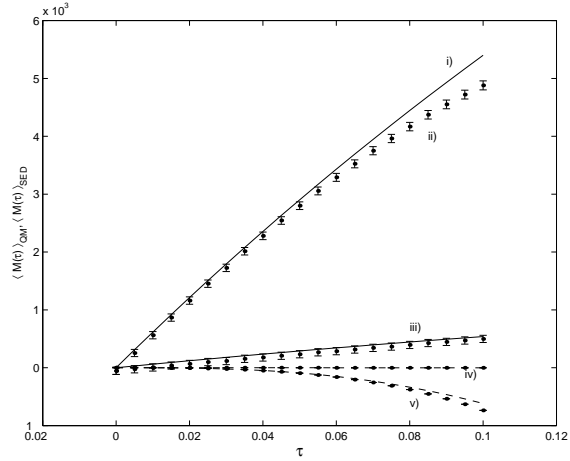,width=75mm}}
\caption{Numerical and analytic results for $\langle {\hat M}(\tau) 
\rangle_{QM}$ 
and $\langle M(\tau) \rangle_{SED}$ for $N=10$, $\gamma=1$ and 
$\cos\Theta=\cos\Phi=1$.
Analytic results for $\langle M(\tau) \rangle_{SED}$ are indicated by solid
lines for i) $g=1$ and iii) $g=0.1$. Numerical results for 
$\langle M(\tau) \rangle_{SED}$ are indicated by dots and with associated error 
bars for ii) $g=1$ and iii) $g=0.1$.
Analytic results for $\langle {\hat M}(\tau) \rangle_{QM}$ are indicated by 
dotted
lines for v) $g=1$ and iv) $g=0.1$. Numerical results for 
$\langle {\hat M}(\tau) \rangle_{QM}$ are indicated by dots with associated 
error bars 
for v) $g=1$ and iv) $g=0.1$.}
\label{fig6}
\end{figure}

\begin{figure}
\center{\epsfig{figure=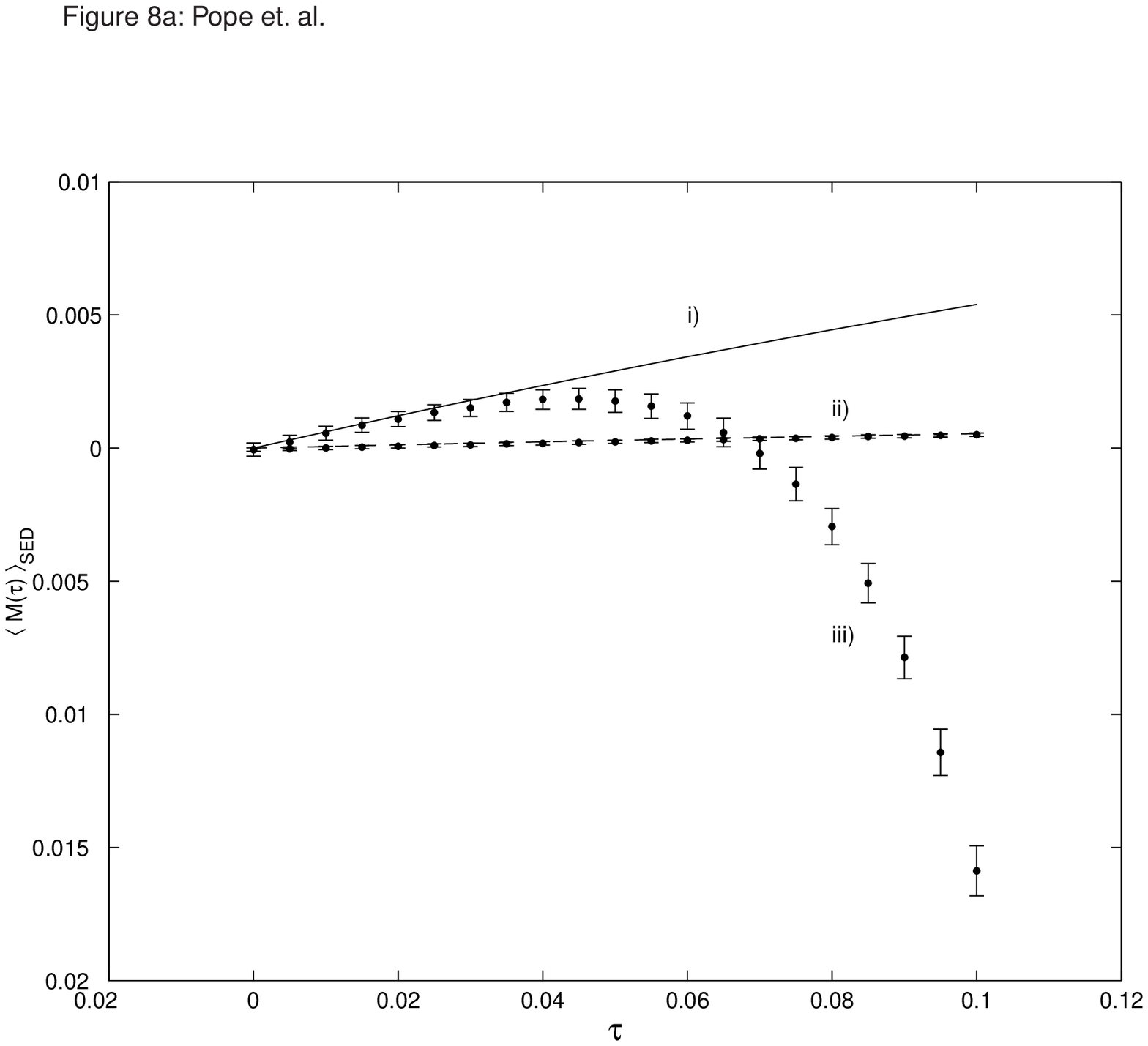,width=75mm}}
\center{\epsfig{figure=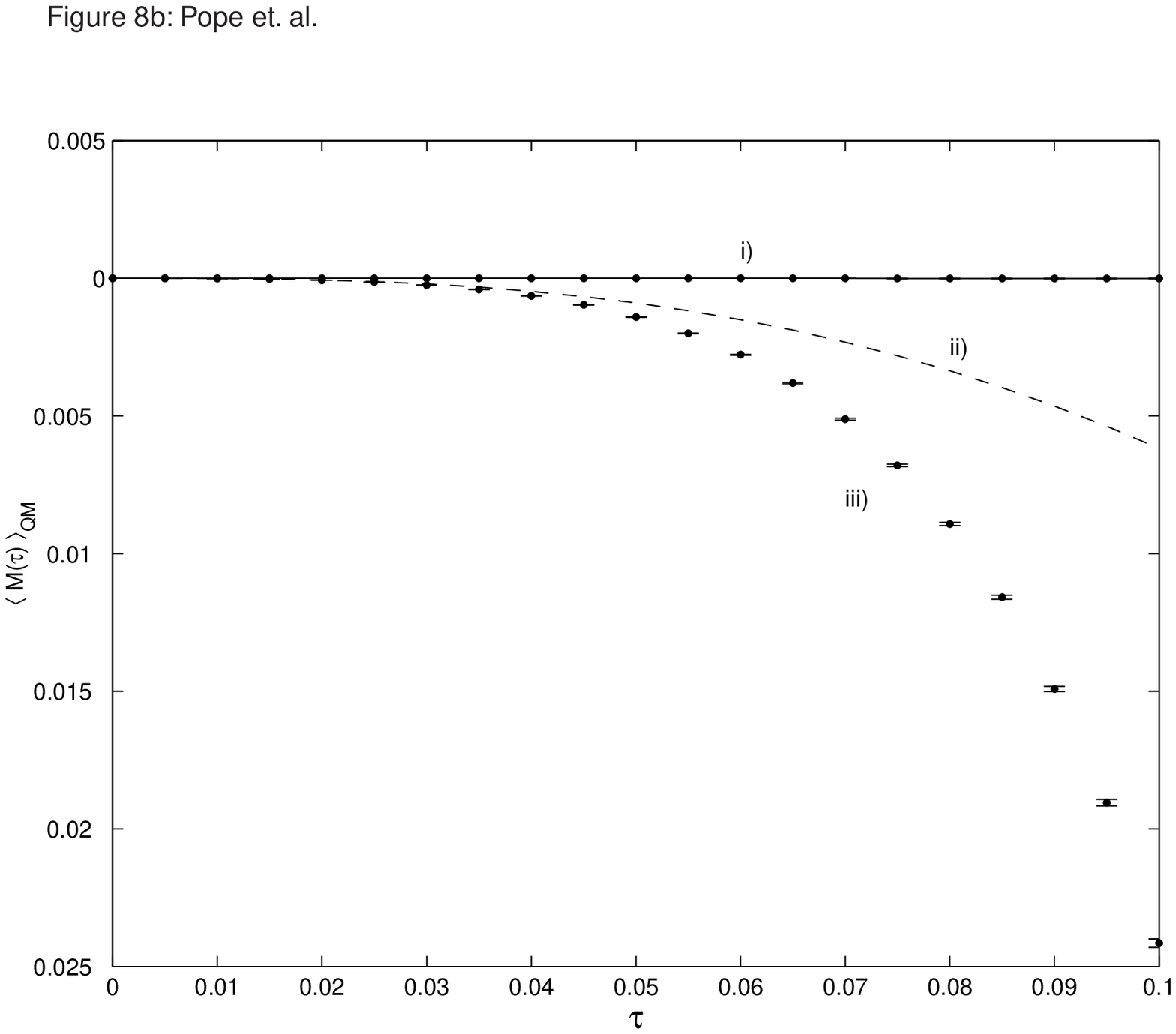,width=75mm}}
\caption{Numerical and analytic results for $\langle {\hat M}(\tau) 
\rangle_{QM}$ 
and $\langle M(\tau) \rangle_{SED}$ for $N=100$, $\gamma=1$ and 
$\cos\Theta=1=\cos\Phi=1$.
(a) Numerical results for $\langle M(\tau) \rangle_{SED}$
represented by dots with associated errorbars 
for ii) $g=0.1$ and iii) $g=1$, and analytic results
for $\langle M(\tau) \rangle_{SED}$ for 
$g=0.1$ (dotted line) and $g=1$ (solid line).
b) Numerical results for $\langle {\hat M}(\tau) \rangle_{QM}$
represented by dots with associated errorbars 
for i) $g=0.1$ and ii) $g=1$, and analytic results
for $\langle {\hat M}(\tau) \rangle_{QM}$ for 
$g=1$ (dotted line) and $g=0.1$ (solid line).}
\label{fig7}
\end{figure}

Results from the numerical and analytic simulations of $\langle M(\tau) 
\rangle_{SED}$ and 
$\langle {\hat M}(\tau) \rangle_{QM}$ over a range of 
$g$ and $N$ values, where $N$ is the average initial number of pump photons
($N=|\epsilon^{2}|$), 
are shown in
Figs~\ref{fig5}-\ref{fig7}. 
In all cases $\theta_{1}=\theta_{2}=\theta_{3}=0$
and relative numerical errors are small.
All $g=0.1$ analytic results are in agreement with their numerical counterparts.
However,
$g=1$ analytic results for N=1 and N=10 are not. 
This disagreement is explained by noting that the analytic results are only 
necessarily valid when $g\ll 1$. 

A number of qualitative trends can be seen in Figs~\ref{fig5}-\ref{fig7}. In 
Figs~\ref{fig5} and \ref{fig6}   
(N=1 and N=10)
the results of SED and quantum mechanics are so distinct that 
they have different signs, with those of quantum mechanics being negative
and those of SED being positive. This trend only holds for short times 
($\tau < 0.07$) in
Figs~\ref{fig7}(a) and (b) 
(N=100). For longer times, SED and quantum mechanics predict the same sign.
This trait is consistent with the fact that Figs~\ref{fig7}(a) and (b) show 
results for the largest number of photons in the pump mode.
SED and quantum mechanics are at their most classical level for this case
and thus might be expected to differ the least. 
For constant N Figs~\ref{fig5}-\ref{fig7} also show that
as $g$ is decreased the results of quantum mechanics and SED become more 
similar.
This occurs because lower $g$ values are associated with larger damping to
nonlinear coupling ratios and therefore move SED and quantum mechanics 
closer to 
the classical domain.

\section{\bf External moments} \label{external_mmt}

Thus far, only intracavity fields have been considered. However,
it is the external fields that leak out of a cavity
that are observed. In realistic systems, intracavity photons are transmitted
through imperfect mirrors into the external environment where they are
detected. Thus, an 
external field analogue of
$\langle M(\tau)\rangle$, $\langle M^{(E)}(\tau_{s},\tau_{f})\rangle$,
where $\tau_{s}$ and $\tau_{f}$ are initial and final 
measurement times, is calculated according to quantum mechanics and SED
to consider what is actually observed in the laboratory.

\begin{figure}
\center{\epsfig{figure=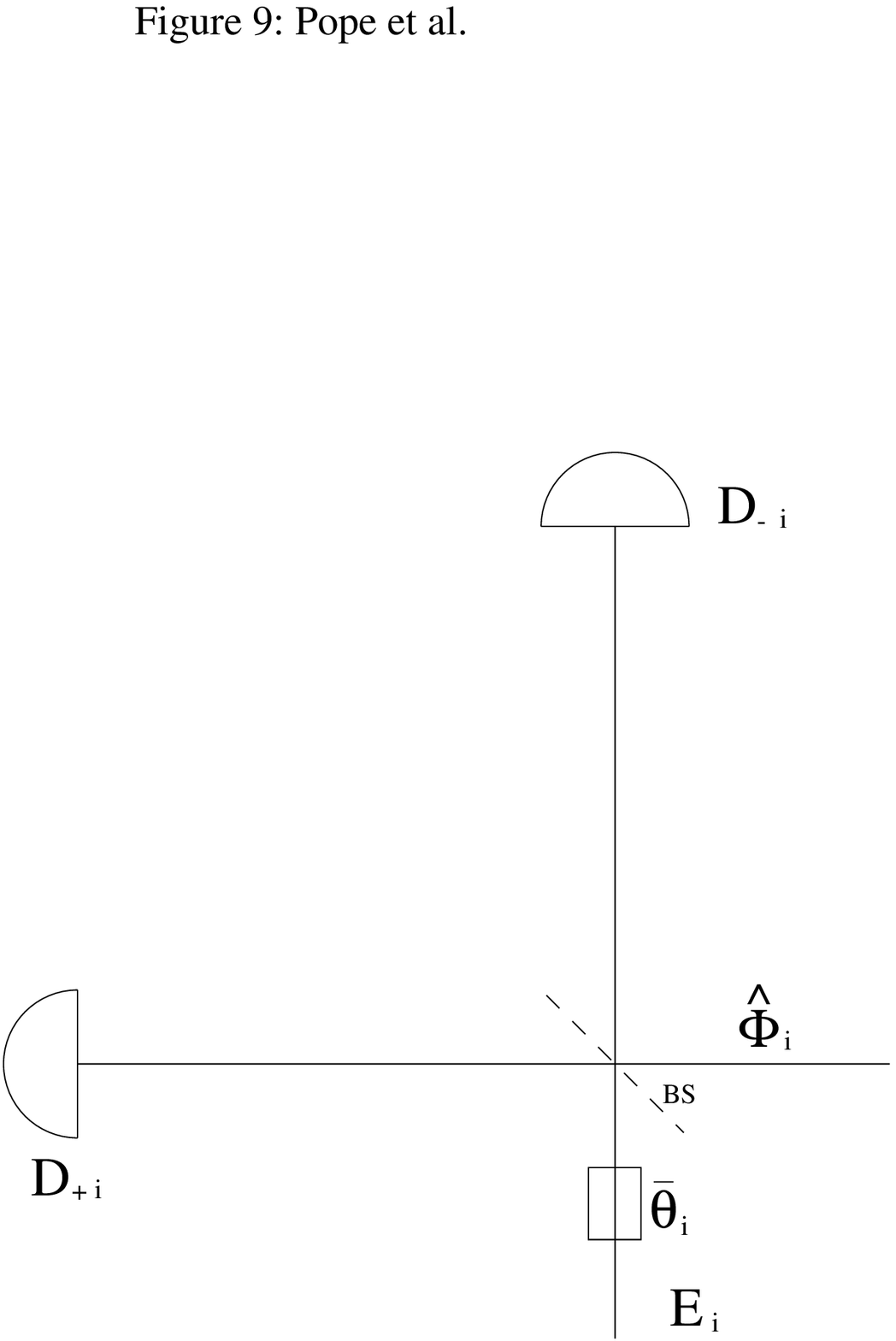,width=75mm}}
\caption{Schematic diagram for balanced homodyne detection. Photodetectors are 
labeled by
$D_{+\:i}$ and $D_{-\:i}$, 
for $i=1,2,3$,
BS is a beamsplitter, ${\bar \theta}_{i}$ is a local oscillator phase variable,
$E_{i}$ is a local oscillator amplitude and ${\hat \Phi}_{i}$ is an 
external signal field operator.}
\label{fig_homo}
\end{figure}

The first step in calculating the external moment $\langle {\hat
M}^{(E)}(\tau)\rangle_{QM}$
for quantum mechanics is defining the external quadrature phase amplitudes 
constituting it.
This is done within the the context
of homodyne detection as quadrature phase amplitudes are commonly measured 
using it.
A schematic diagram for balanced homodyne detection is shown in
Fig.~\ref{fig_homo}.
An external signal field flux ${\hat \Phi}_{i\:OUT}$, where $i=1,2,3$, 
and a local oscillator 
field flux $E_{i}$ are incident on a 50-50 beam splitter BS. 
An external local oscillator phase variable is represented by 
${\bar \theta_{i}}$.
The two field fluxes combine and are detected by two photodiodes $D_{+\:i}$ 
and $D_{-\:i}$. 
The detected photocurrents are then converted to amplified 
electrical currents whose difference is found. An
external quadrature phase amplitude for quantum mechanics
${\hat X}_{i,{\bar \theta_{i}}}^{(E)}$
is defined as this difference yielding, when $E_{i}$ is real,  
\begin{equation}
{\hat X}_{i,{\bar \theta_{i}}}^{(E)}(\tau)=
\frac{e A \eta_{i} E_{i}({\hat \Phi}_{i\:OUT}(\tau) e^{-i{\bar \theta_{i}}}+
{\hat \Phi_{i\:OUT}}^{\dag}(\tau) e^{i{\bar \theta_{i}}})}{2},
\label{ext_quad}
\end{equation}
where e is the magnitude of the charge of an electron, A is an 
amplification factor and $\eta_{i}$ is a detector efficiency factor for both 
detectors
associated with external field modes denoted by $i$.
In realistic experiments detection occurs over
a finite period of time
and thus 
\begin{equation}
\int_{\tau=\tau_{s}}^{\tau=\tau_{f}} \frac{d\tau}{\Gamma} 
{\hat X}_{i,{\bar \theta_{i}}}^{(E)}
(\tau)
\end{equation} 
corresponds to what is observed.
Only the ${\bar \theta_{i}}=0$ case is considered.
Thus, an external moment analogue of the intracavity
moment $\langle {\hat M}(\tau)\rangle_{QM}$ can be defined as
\begin{eqnarray} \label{external}
\langle {\hat M}^{(E)}(\tau_{s},\tau_{f}) \rangle_{QM}&=&\langle 
\Gamma^{-3} \prod_{i=1}^{3} 
\int_{\tau_{i}=\tau_{s}}^{\tau_{i}=\tau_{f}} d\tau_{i} 
\Delta {\hat X}_{i,{\bar \theta_{i}}}^{(E)}(\tau_{i}) \rangle \\ \nonumber
&=& \Gamma^{-3} \int_{\tau_{1}=\tau_{s}}^{\tau_{1}=\tau_{f}} 
\int_{\tau_{2}=\tau_{s}}^{\tau_{2}=\tau_{f}}
\int_{\tau_{3}=\tau_{s}}^{\tau_{3}=\tau_{f}}
\langle \prod_{i=1}^{3} \Delta {\hat X}_{i,{\bar \theta_{i}}}^{(E)}(\tau_{i})
\rangle.   
\end{eqnarray} 

To calculate $\langle {\hat M}^{(E)}(\tau_{s},\tau_{f}) \rangle_{QM}$, the 
relation between 
the unknown external output fields that define it
and known intracavity fields needs to be ascertained. 
Gardiner and Collett \cite{gardiner_collett} have formulated an 
input-output theory
which relates the two via the equation
\begin{equation}
{\hat \Phi}_{i\:OUT}(\tau)=\sqrt{2\Gamma} {\hat a}_{i}(\tau) + 
{\hat \Phi}_{i\:IN}(\tau),
\label{in_out}
\end{equation} 
where ${\hat \Phi}_{i\:IN}(\tau)$ is the input field flux associated with 
intracavity 
mode $i$. All input fields are assumed to be in vacuum states. This allows 
the use of Eq.~(5.3) from \cite{gardiner_collett}, which can be 
expressed as, in this paper's notation,
\begin{eqnarray} \label{vacuum}
\langle {\hat \Phi}^{\dag}_{i\:OUT}(\tau_{1}) {\hat \Phi}^{\dag}_{i\:OUT}
(\tau_{2})... 
{\hat \Phi}^{\dag}_{i\:OUT}(\tau_{n}) {\hat \Phi}_{i\:OUT}(\tau_{n+1}')... 
{\hat \Phi}_{i\:OUT}(\tau_{m}') \rangle & & \\ \nonumber
=(2 \Gamma)^{m/2} \langle \tilde{T} [ {\hat a}^{\dag}_{i}(\tau_{1}) 
{\hat a}^{\dag}_{i}(\tau_{2})...{\hat a}^{\dag}_{i}(\tau_{n}) ]
T[{\hat a}_{i}(\tau_{n+1}')...{\hat a}_{i}(\tau_{m}')]\rangle & &,
\end{eqnarray} 
where $\tilde{T}$ and $T$ are time anti-ordering and time ordering operators
respectively.
Using Eq.~(\ref{ext_quad}), the integrand of Eq.~(\ref{external})
can be expressed in terms of ${\hat \Phi}_{i\:OUT}$ 
and ${\hat \Phi}^{\dag}_{i\:OUT}$. It can then be expressed in
terms of particular ${\hat a_{i}}(\tau_{i})$
and ${\hat a_{i}}^{\dag}(\tau_{i})$ averages using Eq.~(\ref{vacuum}).
In turn, these averages are equivalent to the eight positive-P averages
of the form $\langle \Delta {\cal A}_{1}(\tau_{1}) \Delta {\cal A}_{2}
(\tau_{2})
\Delta {\cal A}_{3}(\tau_{3})\rangle$, where ${\cal A}_{i}$ is either 
$\alpha_{i}$ or $\alpha_{i}^{+}$. As was determined in Section
\ref{stoch_section},
two of these averages,
$\langle \Delta \alpha_{1}(\tau_{1}) \Delta \alpha_{2}(\tau_{2}) 
\Delta \alpha_{3}(\tau_{3}) \rangle$ 
and $\langle \Delta \alpha_{1}^{+}(\tau_{1}) \Delta \alpha_{2}^{+}(\tau_{2}) 
\Delta \alpha_{3}^{+}(\tau_{3}) \rangle$, are of lower order in $g$ than the 
others
and hence dominate when $g \ll 1$.
Thus, the 
external field moment of quantum mechanics $\langle {\hat
M}^{(E)}(\tau_{s},\tau_{f})\rangle_{QM}$ can be 
expressed as, when ${\bar \theta_{i}}=0$, where $i=1,2,3$,
\begin{eqnarray} \label{ext_qt_result}
\langle {\hat M}^{(E)}(\tau_{s},\tau_{f})\rangle_{QM} & \simeq & \frac{\sqrt{2}
(e A \eta E)^{3} \Gamma^{-3/2}}{4}
\int_{\tau_{1}=\tau_{s}}^{\tau_{1}=\tau_{f}}
\int_{\tau_{2}=\tau_{s}}^{\tau_{2}=\tau_{f}} 
\int_{\tau_{3}=\tau_{s}}^{\tau_{3}=\tau_{f}}   
d\tau_{1} d\tau_{2}d\tau_{3} \\ \nonumber 
& & \langle \Delta \alpha_{1} (\tau_{1}) \Delta \alpha_{2} (\tau_{2}) 
\Delta \alpha_{3} (\tau_{3}) \rangle +\langle \Delta \alpha_{1}^{+} (\tau_{1})
\Delta \alpha_{2}^{+} (\tau_{2}) \Delta \alpha_{3}^{+} (\tau_{3}) \rangle, 
\end{eqnarray}
where $\eta=\eta_{i}$ and $E=E_{i}$, where $i=1,2,3$.  
To simplify the algebra 
only the $\tau_{s}=0$ case is investigated, so that only the moment 
$\langle {\hat M}^{(E)}(\tau_{f})\rangle_{QM} (\equiv \langle
{\hat M}^{(E)}(0,\tau_{f})\rangle_{QM})$ is considered. 
External fields are 
only considered for small times ($\tau_{f} << 1$), and so, to a given order in 
$g$, $\langle {\hat M}^{(E)}(\tau_{f})\rangle_{QM}$'s
lowest nonzero order term in $\tau_{f}$ dominates. Hence  
$\langle {\hat M}^{(E)}(\tau_{f}) \rangle_{QM}$ can be approximated by its 
lowest nonzero 
order term in both $g$ and $\tau_{f}$.
Thus,
\begin{equation}
\label{qm_out}
\langle {\hat M}^{(E)}(\tau_{f})\rangle_{QM}  \simeq -\frac{\sqrt{2}}{48}g^{3} 
\epsilon^{2} 
(e A \eta E)^{3} \Gamma^{-3/2} \tau_{f}^{6}. 
\end{equation} 

The SED external moment $\langle M^{(E)}(\tau_{f})\rangle_{SED}$
is now calculated. It is given by the same expression as
$\langle {\hat M}(\tau_{f})\rangle_{QM}$, the right hand
side of Eq.~(\ref{external}) (when $\tau_{s}=0$), except that the
quadrature phase amplitude operator ${\hat X}_{i\:{\bar \theta_{i}}}^{(E)}
(\tau)$, 
is replaced by its SED c-number analogue.
This external SED c-number quadrature phase amplitude is defined as,
when $E_{i}$ is real,
\begin{equation}
X_{i,{\bar \theta_{i}}}^{(E)}(\tau)=
\frac{e A \eta_{i} E_{i}(\beta_{i\:OUT}(\tau) e^{-i{\bar \theta_{i}}}+
\beta_{i\:OUT}^{*} (\tau) e^{i{\bar \theta_{i}}})}{2},
\label{sed_ext_quad}
\end{equation} 
where $\beta_{i\:OUT}(\tau)$ is the output field flux associated with the 
intracavity field denoted by $i$. 
In analogy with Eq.~(\ref{in_out}), it is assumed that the
SED input-output relation is
\begin{equation}
\beta_{i\:OUT}(\tau)=\sqrt{2 \Gamma} \beta_{i}(\tau)+\beta_{i\:IN}(\tau),
\label{sed_in_out}
\end{equation}
where $\beta_{i\:IN}(\tau)$ is the input field flux for the intracavity
mode $i$. 
When all input fields are in vacuum states, as is the case, 
$\beta_{i\:IN}(\tau)$ is a Gaussian white noise with a self correlation
characterized by 
\begin{equation}
\langle \beta_{i\:IN}(\tau_{i})\beta_{i\:IN}^{*}(\tau_{i}') \rangle
=\frac{\delta(\tau_{i}-\tau_{i}')}{2\Gamma}.
\label{noise_corrlns}
\end{equation}
A calculation analogous to the quantum mechanical one earlier in this
section can be performed using Eqs~(\ref{sed_ext_quad}) and (\ref{sed_in_out})
to obtain an expression for 
$\langle M^{(E)}(\tau_{f})\rangle_{SED}$ in terms of particular 
intracavity averages. 
When lowest order nonzero approximations to these averages are considered,
the following result is obtained when ${\bar{\theta}}_{i}$ for $i=1,2,3$,
and $g,\tau_{f} \ll 1$,
\begin{equation} \label{sed_outside_result}
\langle M^{(E)}(\tau_{f}) \rangle_{SED} \frac{\sqrt{2}}{16}\simeq g 
\tau_{f}^{4} (e A \eta E)^{3}
\Gamma^{-3/2}.
\end{equation}
Upon comparing Eq.~(\ref{sed_outside_result}) to result of quantum mechanics in 
Eq.~(\ref{qm_out}), it is seen that the leading order term in $g$
in Eq.~(\ref{qm_out}) is $O(g^{3})$ whilst in Eq.~(\ref{sed_outside_result}) 
it is $O(g)$. Hence, as was the case for the intracavity moment, quantum 
mechanics and SED predict
significantly different results for the observable external field moment 
$\langle M^{(E)}(\tau_{f}) \rangle$.

\section{Signal to noise ratio}

In actual experiments, only finite samples of results are obtained, as opposed 
to infinite
ones. Hence, in practice the population means considered thus far are 
estimated from sample means. These sample means fluctuate from sample to sample
and thus have signal to noise
ratios, which are now determined for small times ($\tau_f<<1$). 
This paper focuses on differences between quantum mechanics and SED.
Thus, a calculation is performed of the signal to noise ratio of the 
difference between the two theories' external sample moments. 
First, the noise of the external sample moment in quantum
mechanics is determined. It is then assumed that
the noise of the external sample moment of SED is the same. Noise results are 
combined with the external moment
results of Section~\ref{external_mmt} to produce $S(\tau_{f})$,
the signal to noise ratio of the difference between the two theories' external
sample moments. This quantity $S(\tau_{f})$ is given by
\begin{eqnarray} \label{noise}   
S(\tau_{f})&=&\frac{\langle | \langle m^{(E)}(\tau_{f})\rangle_{SED} 
- \langle {\hat m}^{(E)}(\tau_{f}) \rangle_{QM} | \rangle }
{\sqrt{s^{2}(\langle m^{(E)}(\tau_{f})\rangle_{SED})+
s^{2}(\langle {\hat m}^{(E)}(\tau_{f}) \rangle_{QM})}} \\ \nonumber
&=& \frac{\sqrt{n-1} |\langle M^{(E)}(\tau_{f})\rangle_{SED} 
- \langle {\hat M}^{(E)}(\tau_{f}) \rangle_{QM} |}
{\sqrt{2}\sigma({\hat M}^{(E)}(\tau_{f}))},
\end{eqnarray}
where $n$ is the number of observations in the sample considered, 
$\langle m^{(E)}(\tau_{f})\rangle_{SED}$ 
and
$\langle {\hat m}^{(E)}(\tau_{f})\rangle_{QM}$
are sample averages of $\langle M^{(E)}(\tau_{f}) \rangle$ according to SED and 
quantum 
mechanics respectively, and $s^{2}(A)$ denotes the 
sample variance of A.
The only significant unknown quantity on the right hand side of 
Eq.~(\ref{noise}) is $\sigma({\hat M}^{(E)}(\tau_{f}))$, which is now 
determined.
Expressing $\sigma({\hat M}^{(E)})$ explicitly yields
\begin{equation}
\sigma({\hat M}^{(E)} (\tau_{f}))=\sqrt{ \langle {\hat M}^{(E)} (\tau_{f})^{2}
\rangle_{QM}
-\langle {\hat M}^{(E)} (\tau_{f}) \rangle_{QM}^{2}}.
\end{equation}
The moment $\langle {\hat M}^{(E)} (\tau_{f}) \rangle_{QM}$ was determined in
Section \ref{external_mmt} and
so $\langle {\hat M}^{(E)} (\tau_{f})^{2} \rangle_{QM}$ is now calculated. 
In the calculation that follows only the ${\bar \theta_{i}}=0$, where
$i=1,2,3$, and $g,\:\tau_{f} \ll 1$ cases are considered.

The moment $\langle {\hat M}^{(E)} (\tau_{f})^{2} \rangle_{QM}$ can be 
expressed in
terms of external quadrature phase operators as
\begin{eqnarray} \label{explicit_sigma}
\langle {\hat M}^{(E)} (\tau_{f})^{2} \rangle_{QM} & = & \langle 
[ \Gamma^{-3} \prod_{i=1}^{3} 
\int_{\tau_{i}=0}^{\tau_{i}=\tau_{f}}
d\tau_{i} \Delta {\hat X}_{i}^{(E)}(\tau_{i}) ]^{2} \rangle \\
\nonumber
& = &  \Gamma^{-6} 
\int_{\tau_{1}=0}^{\tau_{1}=\tau_{f}} \int_{\tau_{1}'=0}^{\tau_{1}'=\tau_{f}}
\int_{\tau_{2}=0}^{\tau_{2}=\tau_{f}} \int_{\tau_{2}'=0}^{\tau_{2}'=\tau_{f}}
\int_{\tau_{3}=0}^{\tau_{3}=\tau_{f}} \int_{\tau_{3}'=0}^{\tau_{3}'=\tau_{f}}
d\tau_{1} d\tau_{1}' d\tau_{2} d\tau_{2}' d\tau_{3} d\tau_{3}'
\langle \prod_{i=1}^{3} \Delta {\hat X}_{i}^{(E)}(\tau_{i})  
\Delta {\hat X}_{i}^{(E)}(\tau_{i}') \rangle,
\end{eqnarray}
where ${\hat X}_{i}^{(E)}(\tau_{i})={\hat X}_{i\:{\bar \theta_{i}}=0}^{(E)}
(\tau_{i}).$
The integrand of 
Eq.~(\ref{explicit_sigma}), which is denoted by K, 
can be expressed as 
\begin{equation}
K=\prod_{i=1}^{3}\langle\Delta {\hat X}_{i}^{(E)} (\tau_{i}) 
\Delta {\hat X}_{i}^{(E)}(\tau_{i}') \rangle + f(\tau_{i},\tau_{i}'), 
\end{equation}
where $f(\tau_{i},\tau_{i}')$ is a function which includes terms resulting
from coupling between modes. These coupling terms vanish when $g=0$ and thus 
are at least O(g).
It follows that $K$ can be be re-expressed as 
\begin{equation}
K=\prod_{i=1}^{3} \langle \Delta {\hat X}_{i}^{(E)}(\tau_{i}) 
\Delta {\hat X}_{i}^{(E)}(\tau_{i}')\rangle + O(g).
\label{k_Eq.}
\end{equation}

The moment $\langle \Delta {\hat X}_{i}^{(E)}(\tau_{i}) 
\Delta {\hat X}_{i}^{(E)}(\tau_{i}')\rangle$ is 
now calculated using a normally ordered approach that has been previously
employed to solve similar problems \cite{carmichael},\cite{drummond_two}.
This method expresses  
$\langle \Delta {\hat X}_{i}^{(E)}(\tau_{i}) 
\Delta {\hat X}_{i}^{(E)}(\tau_{i}')\rangle$
in terms of normally ordered
photocurrent averages and then determines these averages. It first
defines ${\hat X}_{i}^{(E)}(\tau_{a})$, where $\tau_{a}$ is any $\tau$ 
variable,
as the difference between the amplified {\em electrical} currents, 
${\hat X}_{+\:i}^{(E)}(\tau_{a})$ and ${\hat X}_{-\:i}^{(E)}(\tau_{a})$,
produced by the 
{\em photo}currents detected at the detectors $D_{+\:i}$ and $D_{-\:i}$ in 
Fig.~\ref{fig_homo}
in Section \ref{external_mmt}.
Using this definition (${\hat X}_{i}^{(E)}(\tau_{a})=
{\hat X}_{+\:i}^{(E)}(\tau_{a})-{\hat X}_{-\:i}^{(E)}(\tau_{a})$), 
$\langle \Delta {\hat X}_{i}^{(E)}(\tau_{i}) 
\Delta {\hat X}_{i}^{(E)}(\tau_{i}')\rangle$ can be expressed as
\begin{equation} \label{expand}
\langle \Delta {\hat X}_{i}^{(E)}(\tau_{i}) 
\Delta {\hat X}_{i}^{(E)}(\tau_{i}')\rangle  = 
\langle(\Delta {\hat X}_{+\:i}^{(E)}(\tau_{i})-\Delta {\hat X}_{-\:i}^{(E)}
(\tau_{i})) 
(\Delta {\hat X}_{+\:i}^{(E)}(\tau_{i}')-\Delta {\hat X}_{-\:i}^{(E)}
(\tau_{i}'))\rangle. 
\end{equation} 
Upon expansion, the right hand side of Eq.~(\ref{expand})
contains two types of terms, those of the form
$\langle {\hat X}_{C\:i}^{(E)} (\tau_{a})\rangle$, 
where C is either + or -,
and those of the form
$\langle {\hat X}_{C\:i}^{(E)} (\tau_{i}) {\hat X}_{D\:i}^{(E)}
(\tau_{i}')\rangle$,
where D is either + or -.
Terms of the form 
$\langle {\hat X}_{C\:i}^{(E)}(\tau_{a})\rangle$ 
are given by the equation 
\begin{equation}
\langle {\hat X}_{C\:i}^{(E)}(\tau_{a})\rangle= 
\int_{-\infty}^{+\infty}
\frac{ds_{1}}{\Gamma} G_{C\:i}^{(1)}(s_{1}) J^{0}(\tau_{a}-s_{1}),
\end{equation}
where $G_{C\:i}^{(1)}(s_{1})$ is a first order Glauber correlation function
and $J^{0}(\tau_{a}-s_{1})$ is an electrical current pulse
produced by a single photodetection event.
In following previous work \cite{carmichael}, square electrical 
current pulses of the form \\
\begin{eqnarray}
J^{0}\left(a-b \right)=\left\{
\begin{array}{ll}
Ae \Gamma/\tau_{d} & b\leq a\leq b+\tau_{d} \\
0 & a<b\;{\rm and}\;a> b+\tau_{d} \\
\end{array}
\right.
\end{eqnarray}
are considered in the limit of $\tau_{d}\rightarrow 0$, which is
taken at some appropriate later stage of the calculation. 
The Glauber correlation function $G_{C\:i}^{(1)}(s_{1})$
can be expressed as a power series
in $g$ and $s_{1}$ and thus as 
$\sum_{m,n=0}^{\infty,\infty} c_{mn} g^{m} {s_{1}^{n}}$.
Due to the form of $J^{0}(\tau_{a}-s_{1})$,
when $\tau_{a}<<1$, as is being assumed,
only photodetection events at small times $s_{1}$, contribute to 
$\langle {\hat X}_{C\:i}^{(E)} (\tau_{a})\rangle$. This fact, coupled
with the knowledge that only the $g << 1$ case is considered, 
means that the $n=m=0$ term 
in the power series for $G^{(1)}_{C\:i}(s_{1})$ dominates when 
$J^{0}(\tau_{a}-s_{1})$ is nonzero. 
Hence, upon calculating this dominant term
by expressing ${\hat \Phi}_{i}$ and ${\hat \Phi}^{\dag}_{i}$
in terms of intracavity field operators,
in the limit of
large local oscillator amplitude,
\begin{equation}
G_{C\:i}^{(1)}(s_{1}) \simeq \frac{\eta_{C\:i}}{2}E_{i}^{2},
\end{equation}
where $\eta_{C\:i}$ is a detector efficiency factor for the photodetector
$D_{C\:i}$.
It follows that 
\begin{equation} \label{one_current}
\langle {\hat X}_{C\:i}^{(E)} (\tau_{a}) \rangle \simeq
\frac{\eta_{C\:i} E^{2}_{i} A e}{2}. 
\end{equation}

Terms of the form 
$\langle {\hat X}_{C\:i}^{(E)}  (\tau_{i}) {\hat X}_{D\:i}^{(E)}  
(\tau_{i}')\rangle$ 
in Eq.~(\ref{expand}) can be expressed as
\begin{eqnarray} \label{two_order}
\langle {\hat X}_{C\:i}^{(E)} (\tau_{i}) {\hat X}_{D\:i}^{(E)} 
(\tau_{i}')\rangle
&=& \delta_{CD} \int_{-\infty}^{+\infty}\frac{ds_{1}}{\Gamma}
G^{(1)}_{C\:i}(s_{1})J^{(0)}
(\tau_{i}-s_{1})J^{(0)}(\tau_{i}'-s_{1}) \\ \nonumber
& & +\int_{-\infty}^{+\infty}\int_{-\infty}^{+\infty}
\frac{ds_{1}ds_{2}}{{\Gamma}^{2}}G^{(2)}_{C,D\:i}(s_{1},s_{2})
J^{(0)}(\tau_{i}-s_{1}) J^{(0)}(\tau_{i}'-s_{2}),
\end{eqnarray}
where $G^{(2)}_{C,D\:i}(s_{1},s_{2})$ is a second order Glauber 
correlation function
and $\delta_{C,D}$ is one when C and D are the same and zero otherwise.
In the limit of $\tau_{d} \rightarrow 0$,
\begin{equation}
\int_{-\infty}^{+\infty}\frac{ds_{1}}{\Gamma} 
G^{(1)}_{C\:i}(s_{1}) J^{(0)}
(\tau_{i}-s_{1})J^{(0)}(\tau_{i}'-s_{1}) \simeq
\frac{(eAE_{i})^{2}\eta_{C\:i} \delta(\tau_{i}-\tau_{i}')\Gamma}{2}
\end{equation}
to leading nonzero order in $g, \tau_{i}$ and $\tau_{i}'$. It is of equal order in
$g$ and 
lower order in 
$\tau_{i}$ and $\tau_{i}'$ than the second term in Eq.~(\ref{two_order})
and hence is much larger than this second term 
when it is nonzero as the $\tau_{i},\tau_{i}'<<1$ case is being considered. 
Thus
\begin{equation} \label{delta_lead}
\langle {\hat X}_{C\:i}^{(E)} (\tau_{i}) {\hat X}_{D\:i}^{(E)} (\tau_{i}')
\rangle \simeq 
\frac{\delta_{CD}(AeE_{i})^{2} \eta_{C\:i} \delta(\tau_{i}-\tau_{i}')\Gamma}
{2}.
\end{equation}

From Eqs~(\ref{one_current}) and (\ref{delta_lead}) it can be seen that 
the single integral terms in 
$\langle {\hat X}_{+\:i}^{(E)} (\tau_{i}) 
{\hat X}_{+\:i}^{(E)} (\tau_{i}')\rangle$ and 
$\langle {\hat X}_{-\:i}^{(E)} (\tau_{i}) 
{\hat X}_{-\:i}^{(E)} (\tau_{i}')\rangle$
are of the same order in $g$ and lower order in $\tau_{i}$ and $\tau_{i}'$ than 
any other terms
contributing to $\langle {\hat X}_{i}^{(E)} (\tau_{i}) 
{\hat X}_{i}^{(E)} (\tau_{i}')\rangle$ and hence dominate.
It follows that
\begin{equation}
\langle {\hat X}_{i}^{(E)} (\tau_{i}) {\hat X}_{i}^{(E)} 
(\tau_{i}')\rangle \simeq (AeE)^{2} \eta_{i} \delta(\tau_{i}-\tau_{i}') \Gamma,
\label{leading}
\end{equation}
where $\eta_{i}=\eta_{C\:i}=\eta_{D\:i}$ and $E=E_{i}$, where $i=1,2,3$. 
As right hand side of Eq.~(\ref{leading}) is  
$O(g^{0})$, $\prod_{i=1}^{3} \langle \Delta {\hat X}_{i}^{(E)}(\tau_{i}) 
\Delta {\hat X}_{i}^{(E)}(\tau_{i}')\rangle$ is also $O(g^{0})$
and hence from Eq.~(\ref{k_Eq.}), 
\begin{equation}
K \simeq ((AeE)^{2} \Gamma)^{3} \prod_{i=1}^{3} \eta_{i}
\delta(\tau_{i}-\tau_{i}'). 
\end{equation}
Substituting this approximation for K into Eq.~(\ref{explicit_sigma}) yields
\begin{eqnarray}
\langle {\hat M}^{(E)} (\tau_{f})^{2} \rangle & \simeq & 
\Gamma^{-6} ( (eAE)^{2} \eta \Gamma)^{3}  
\prod_{i=1}^{3} \int_{\tau_{i}=0}^{\tau_{i}=\tau_{f}}
\int_{\tau_{i}'}^{\tau_{i}'=\tau_{f}} 
d\tau_{i} d\tau_{i}' \delta(\tau_{i}-\tau_{i}') \\ \nonumber
&=& \left[ \frac{(eAE)^{2} \eta \tau_{f}}{\Gamma} \right]^{3},
\end{eqnarray}
where $\eta=\eta_{i}$, where $i=1,2,3$.
Thus 
\begin{equation}
\sigma({\hat M}^{(E)}(\tau_{f})) \simeq \bigl[\frac{(eAE)^{2} \eta \tau_{f}}
{\Gamma} \bigr]^{3/2}.
\end{equation}
Hence, the signal to noise ratio of the difference between the external 
sample moments
of quantum mechanics and SED is 
\begin{equation}
S(\tau_{f}) \simeq \frac{\sqrt{n-1} \eta^{3/2} g \tau_{f}^{5/2}}{16}.
\end{equation}

\section{Realistic systems} \label{real}

Realistic parameter values are now considered to determine if the theoretical
difference between SED and quantum mechanics could be observed experimentally.
In particular, the signal to noise ratio of the difference between the sample
moments of quantum mechanics and SED $S$ is calculated using realistic 
parameter values  
for nondegenerate parametric oscillators containing the 
commonly used crystals, silver gallium selinide (${\rm AgGaSe_{2}}$) and
potassium titanyl phosphate (KTP).  
The non-linear interaction strength G for parametric down conversion is given 
by \cite{realistic}
\begin{equation}
G\simeq d_{eff}\sqrt{\frac{2 \hbar \omega_{1} \omega_{2} \omega_{3}}
{\epsilon_{0}V}}\frac{l}{L},
\end{equation}
where V is the cavity volume, l is the crystal length and L the cavity
length. Cavity and crystal length values of 10cm are chosen.
The cavity volume V is given by the formula $V=\pi \Omega^{2} L$,
where $\Omega$ is the spot size. This volume is minimized
in order to maximize G and thus the external difference between 
quantum mechanics and
SED. It is assumed that the damping constant used to scale time
$\Gamma$ equals the unscaled damping constant for each mode $\Gamma_{i}$ 
($\Gamma
=\Gamma_{i}$). This common damping constant $\Gamma$  
is calculated from the formula
$\Gamma=T \times \frac{c}{2L}$, where c is the speed of light and
T is a mirror transmission coefficient. A $T$ value of $T=0.01$ is used. 
Using the above information, Table~\ref{table_1} shows realistic parameter
values for $d_{eff}$, $V$, $G$, pump, signal and idler wavelengths, and
resulting $g$ and $\Gamma$ values.
Results for $\langle {\hat M}^{(E)}\rangle_{QM}$ and 
$\langle M^{(E)}\rangle_{SED}$ are
obtained using Eqs~(\ref{qm_out}) and (\ref{sed_outside_result}) for when 
$\eta=1$, $\tau_{f}=0.1$, $E=10^{9}s^{-1/2}$, $A=1/e$ and $\epsilon=10^{3}$.
These are displayed in Table~\ref{table_2}, 
which shows that the external results of quantum mechanics
and SED differ greatly. 
Due to local oscillator amplification, they are also macroscopically distinct
with respect to photon number, even though the initial number of
intracavity photons is small on average.
Another appealing feature of the difference between the two theories is that
detector efficiencies approach one as photodiodes as opposed to
photomultipliers are used for detection. Thus, no fair sampling assumptions 
need to be made.

\begin{figure}
\center{ \epsfig{figure=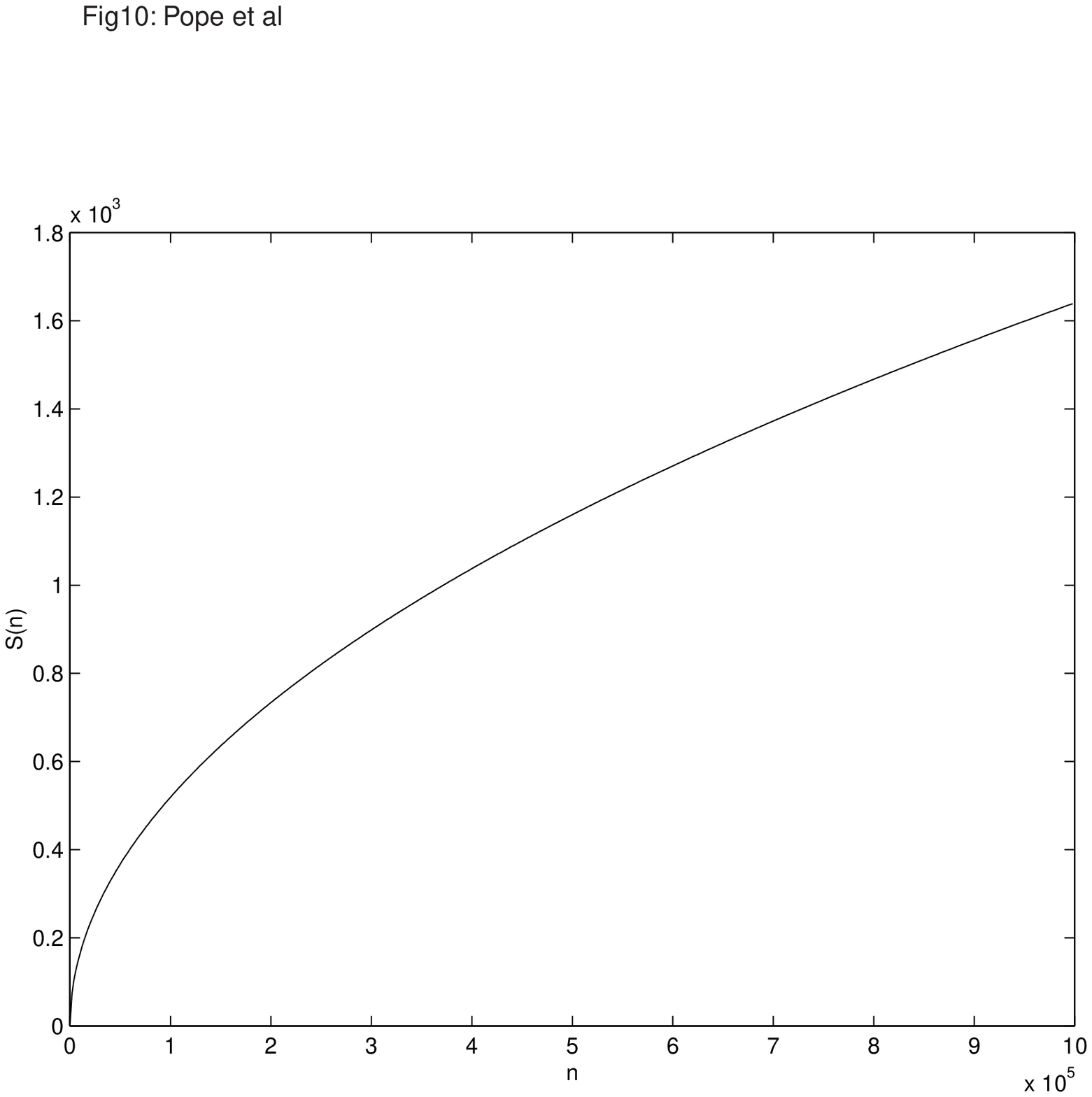,width=75mm}}
\caption{Plot of signal to noise ratio of the difference between the sample
moments of quantum mechanics and SED
S versus sample size n for ${\rm AgGaSe_{2}}$ for g=$8.3 \times 10^{-3}$, 
$\tau_{f}=0.1$ and $\eta=1$.}
\label{snr_ag}
\end{figure}

\begin{figure}
\center{ \epsfig{figure=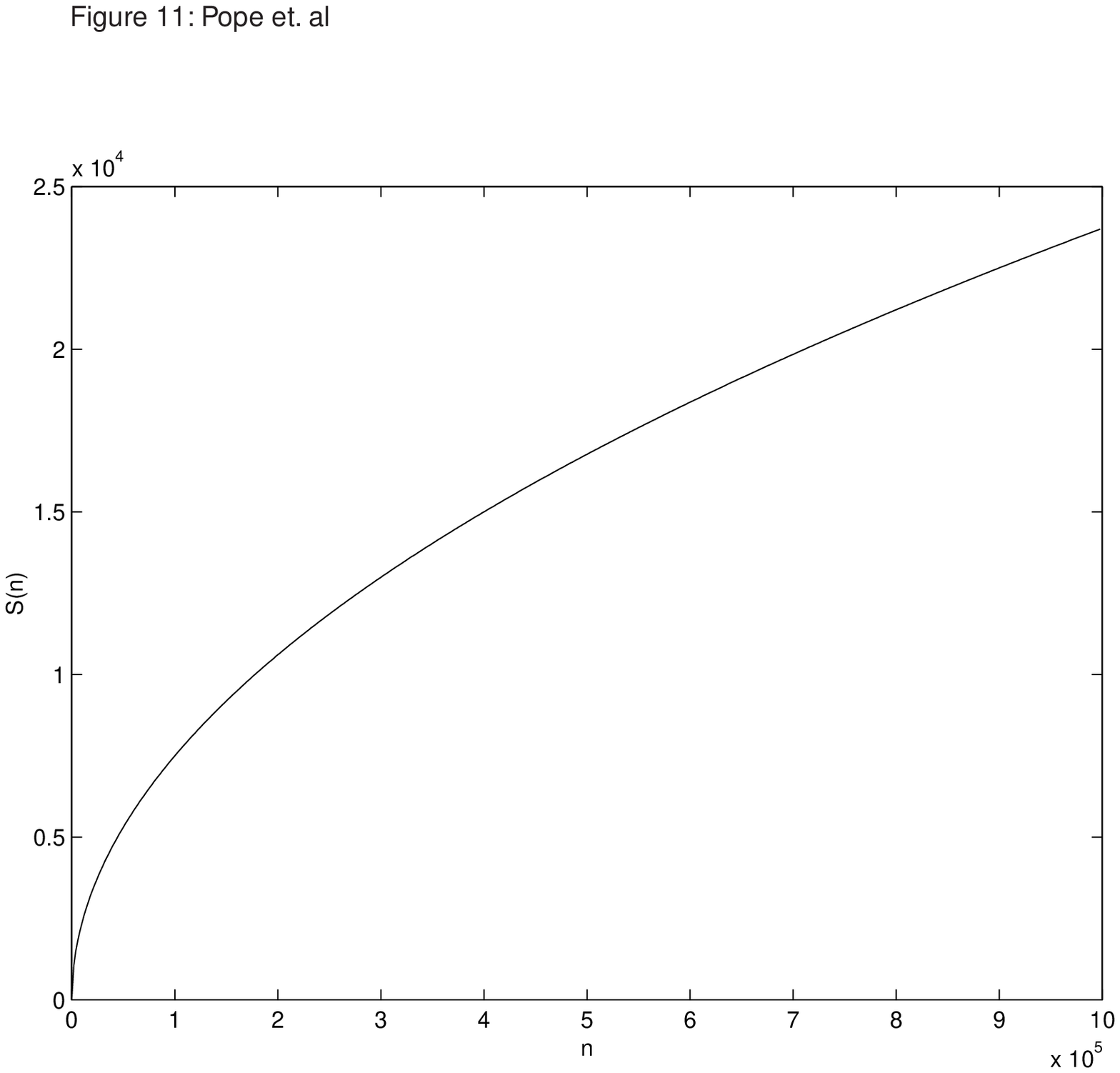,width=75mm}}
\caption{Plot of signal to noise ratio of the difference between the sample
moments of quantum mechanics and SED
S versus sample size n for KTP for g=$1.2 \times 10^{-3}$,
$\tau_{f}=0.1$ and $\eta=1$.}
\label{snr_ktp}
\end{figure}

The question remains of whether or not the population difference between
SED and quantum mechanics could be reliably observed in a finite sample of
results. To answer it, $S$ is now considered.
Figs~\ref{snr_ag} and \ref{snr_ktp} show graphs of $S$
versus sample size $n$ for ${\rm AgGaSe_{2}}$ and KTP for the same parameter
values as used in the last paragraph.
These show reasonable $S$ values and indicate that 
large sample sizes must be obtained to produce a signal to noise
ratio of one, the smallest signal to noise ratio required to clearly observe 
the signal. In particular, sample sizes of $1.8 \times 10^{13}$ (KTP) and 
$3.7 \times 10^{11}$ (${\rm AgGaSe_{2}}$) need to be obtained to generate a 
signal to noise ratio of one. An individual observation takes a time of the 
order 
$t=\tau_{f}/\Gamma=6.7 \times 10^{-9}\:s$ and so, 
assuming minimal time delay between measurements, 
$1.8 \times 10^{13}$ observations would take about 
33 hours and $3.7 \times 10^{11}$ observations about 41 minutes.  
It is conceivable that measurements could be taken over both times.
Furthermore, as the signal to noise ratio scales as 
$\frac{1}{g}$, higher $g$ materials would enable the difference to be observed 
even more readily.

\section{discussion}

It has been shown that there exist a significant, potentially experimentally 
observable,
difference between quantum mechanics and SED. Due to local oscillator
amplification, this difference can involve macroscopically 
distinct external fields for the two theories. Thus, it can be considered 
macroscopic if it is legitimate to include the local oscillators as part of the
system and not as external measuring apparatuses. The difference is also
potentially 
experimentally observable, as a realistic system and state are considered
and is present at realistic parameter values. 
The system is practical as parametric oscillators and balanced homodyne 
detection
are widely used, and damping is included.
The state is realistic as the initial intracavity coherent state can be  
approximated well by a laser. It follows that the
difference can be seen as providing the basis for an 
{\em experimentally achievable macroscopic} test of quantum mechanics against 
one local hidden variable theory (SED). Such a test is significant as all 
experimental tests of quantum mechanics against local hidden variable theories
to date have been microscopic. It is true that many macroscopic tests have been 
proposed, but most of them consider highly idealized states or systems that are 
not currently able to be experimentally implemented. 
In particular, many of them do not consider damping, even though it is known to
rapidly destroy the correlations of quantum mechanics present in 
Schroedinger cat \cite{cat} and other entangled states.
The calculations in this paper do include damping and show that the 
difference between SED and quantum mechanics is not overly sensitive to it.
Most importantly, it remains for realistic damping values. The
test proposed in this paper can be seen as being in the novel and largely 
unexplored domain of macroscopic experimental tests of quantum mechanics.
 
Even if the local oscillators are not included as part of the
system investigated, the external difference between quantum mechanics and SED 
is still at least mesoscopic as average initial pump photon numbers up to 
$10^{6}$ are considered.
From this perspective, the difference is still distinct from many earlier 
microscopic
ones known to exist between quantum mechanics and all local hidden variable 
theories.
It is also, perhaps, more surprising than some of them as it occurs 
in a larger particle number system.

Two noteworthy features of the external difference between 
quantum mechanics and SED are
that it involves continuous variables and high efficiency detection. That it
involves continuous variables is significant because most previous differences
between quantum mechanics and local hidden variable theories have involved
discrete ones. Furthermore, it is, perhaps, more surprising that a
difference between quantum mechanics and a local hidden variable theory can be
found for continuous variables as continuous variables are more closely related 
to classical
ones (which are all continuous) than discrete ones. Low detector
efficiency forms the basis of a significant loophole in most tests 
between quantum mechanics and SED to date \cite{loophole}. The use of 
photodiodes for detection
in the scheme discussed means that such a loophole is avoided.

The calculations in Section \ref{real} show it is difficult to observe the
external difference between quantum mechanics and SED. This is mainly a result 
of small
experimental nonlinearities. They cause few signal and idler photons to be 
created and thus
the experimental signal is weak relative to its noise. For small enough
measurement samples, SED results cannot be clearly distinguished from those of 
quantum mechanics. This fact is
consistent with the knowledge that SED reproduces many features of quantum
mechanics. However, it is a distinct theory and does differ from quantum
mechanics in particular cases, as this paper has shown.

The external difference between quantum mechanics and SED would be easier to 
observe if
larger nonlinear coupling constants were used.
These could be achieved by using organic nonlinear 
crystals such as 
{\em N}-(4-nitrophenyl)-{\em L}-prolinol (NPP) \cite{organic}. However, 
phase matching would be difficult with such crystals. In addition,
they are typically only transparent within a small frequency range. 
Alternatively, higher nonlinearities could be achieved by using 
Josephson-parametric amplifiers \cite{josephson}, 
which can have even larger nonlinearities than organic nonlinear crystals. 
Another possibility, in the area of atom optics, is to to
utilize BEC nonlinear effects, in which atom-molecule coupling is induced
through photon-associaton \cite{BEC}.

To conclude, this paper compared particular moments of 
quantum mechanics to those of SED for the nondegenerate parametric 
oscillator. Both internal and external moments 
were considered and an analytic iterative technique showed them both to be 
cubic in the system's nonlinear coupling constant for 
quantum mechanics and linear for SED. Numerical simulations were performed to 
check the approximate intracavity analytic result and were in agreement with 
them when the system's nonlinear coupling constant was much less than one.
Realistic parameter values were considered and it was shown
that the external sample difference between SED and quantum mechanics 
had a small signal to noise ratio in typical parametric oscillators. The 
presence of intense local oscillators means that the results could be seen as
providing the basis for a macroscopic experimental test of quantum mechanics 
against SED.
\section*{acknowledgements} 

DTP would like to thank Professor Gerard Milburn, Dr
Howard Wiseman, Dr Karen Kheruntsyan, Professor Brian Orr, Michael Gagen
and Cynthia Freeman for their assistance with the paper. He would also like to 
thank The University of
Queensland for its financial support. WJM acknowledges support from the 
Australian Research Council.

\newpage

\newpage

\begin{table} 
\begin{tabular} {ccccccc}
CRYSTAL & $d_{eff}(pmV^{-1})$ & $V(m^{3})$ & $\lambda_{3} \simeq 
\lambda_{2}/2 \simeq \lambda_{1}/2 (\mu m)$ 
& $G(s^{-1})$ & $
\Gamma(s^{-1})$ & $g(=\frac{G}{\Gamma}$) \\ \tableline
${\rm AgGaSe_{2}}$ & $33\:(at\:\lambda=2.1\mu m$) \cite{para_book} 
& $1.0 \times 10^{-9}$& $1.4$ 
& $1.3 \times 10^{5}$ & $1.5\times10^{7}$ & $8.3 \times 10^{-3}$ \\ 
KTP & $7.2\:(at\:\lambda=2.3\mu m$) \cite{para_book} 
& $1.6 \times 10^{-9}$ & $1.6$ 
& $1.8 \times 10^{4}$ & $1.5\times10^{7}$ & 
$ 1.2 \times 10^{-3}$ \\
\end{tabular}
\caption{Table of realistic values for the parameters $d_{eff}$, volume V,
pump, signal and idler
wavelengths $\lambda_{3},\lambda_{2}$ and $\lambda_{1}$, nonlinear
coupling constant G, damping
constant $\Gamma$ and nonlinear coupling constant to damping ratio $g$
for the nonlinear crystals
${\rm AgGaSe_{2}}$ and KTP.}
\label{table_1}
\end{table}

\newpage

\begin{table} 
\begin{tabular} {ccc}
CRYSTAL & $\langle {\hat M^{(E)}}(\tau_{f})\rangle$ & $\langle M^{(E)}(\tau_{f})
\rangle_{SED}$ 
\\ \tableline
${\rm AgGaSe_{2}}$ & -$2.9 \times 10^{8}$ & $1.3 \times 10^{9}$ \\ 
KTP & $-8.8 \times 10^{5}$ & $1.8 \times 10^{8}$ \\
\end{tabular}
\caption{Table of the external moments of  
quantum mechanics and SED for
${\rm AgGaSe_{2}}$ and KTP.}
\label{table_2}
\end{table}

\end{document}